# Operando Characterization of Volume Changes in Lithium-Ion Battery Electrodes during Cycling using Isotope Multilayers

Erwin Hüger[1,2,*], Daniel Uxa[1], Lars Dörrer[1,2], Jochen Stahn[3], and Harald Schmidt[1,2]

[1]Clausthal University of Technology, Solid State Kinetics Group, Institute of Metallurgy, 38678 Clausthal-Zellerfeld, Germany.
[2]Clausthal Center for Materials Technology, 38678 Clausthal-Zellerfeld, Germany.
[3]Center for Neutron and Muon Sciences, Paul Scherer Institute, Villigen PSI 5232, Switzerland

*Corresponding author: erwin.hueger@tu-clausthal.de

**Abstract**

This study reports on advancements in operando characterization of volume changes in lithium-ion battery (LIB) electrode materials during electrochemical cycling. Volume changes are crucial for LIB operation because they are related to the amount of stored energy as well as LIB integrity, performance, and safety. The study introduces a method based on isotope multilayers as active material to track the intrinsic modification of electrode volume in real time under operating conditions with operando neutron reflectometry. A $^{nat}$Ge/$^{73}$Ge multilayer film is used as a model system to measure the volume change of amorphous germanium electrodes during charging and discharging. Isotope modulation produces a Bragg peak in the neutron reflectivity pattern, sensitive only to the modification of volume within the active material of the electrode. Battery side reactions, such as the growth and reduction of the solid-electrolyte interphase is excluded. Using this method, the volume modification as a function of Li content x in Li$_x$Ge can easily be derived from the scattering vector position of the Bragg peak without fitting numerous complex reflectivity patterns. The experiments show a reversible volume change of amorphous germanium of up to 250 % for x ≈ 3, which appears to be largely independent of current density, cycle number, and the thickness of the individual Ge layers. Also, there are tentative indications that the crystallization and re-amorphization of Li$_x$Ge are not influencing the volume change.

**Keywords**: neutron reflectometry, lithium-ion batteries, germanium electrodes, volume change, isotope multilayers.

# 1. Introduction

In order to develop a deeper understanding of functional energy materials, it is necessary to observe their performance under real conditions. Advances in analytical techniques allow for measurements to be recorded during operation [1,2], known as operando measurements. Overcoming challenges in electrochemical energy storage depends on further developing and improving these technologies [1]. The present work contributes to this effort.

Unlike other methods, neutron beams can penetrate deeply into materials, enabling the operando investigation of devices (e.g., electrochemical cells [3-16] or furnaces [17-23]). In addition, neutrons are sensitive to light elements (e.g. Li) and allow for contrast variation by isotope substitution [15,17-27]. The latter is precisely what is used in the current work to enable the selective measurement of volume changes inside high-capacity electrode active materials of lithium-ion batteries (LIBs).

Neutron reflectometry (NR) is a non-destructive scattering technique that enables the determination of depth profiles of neutron scattering length density (nSLD), allowing detailed analysis of the structure and composition of thin films and layered materials [2-27]. Experiments on in situ examination of corrosion and corrosion protection [2], (ii) ex situ [26,27] and in situ [17–21,23] determination of self-diffusivities, (iii) in situ [22,24] measurement of lithium permeation through LIB relevant materials, and (iv) in situ [13,16] and operando [3–8, 10–12] investigation of LIB operation were carried out. Various systems were investigated concerning the last three topics, including self-diffusion in crystalline [27] and amorphous silicon films [23], crystalline [26] and amorphous germanium films [18-20], and various amorphous silicon-germanium compounds [17,21]. Ex situ NR was also applied to examine Li self-diffusivity in crystalline $LiNbO_3$, $Li_3NbO_4$, and $LiAlO_2$, as well as in their amorphous film counterparts [25]. Li permeability was measured in situ with NR through thin amorphous silicon films in contact with lithium-metal-oxide layers [22, 24]. The behavior of LIB electrode active materials during LIB operation was investigated for crystalline [6] and amorphous silicon [3-5, 8-13,16], amorphous carbon [7], and amorphous germanium [15]. Reviews focusing on the application of NR to the study of LIB materials can be found in references [9,14,25]. NR has primarily been used to characterize the solid-electrolyte interphase (SEI) [7,9,13,14,28,29]. This layer forms at the surface of LIB electrodes, such as carbon [7,28,29] and silicon [13,29-31] during cycling. NR experiments preserve the integrity, configuration, and operational functionality of the devices under investigation, as also demonstrated in the current study. As above mentioned, a recent study describes NR as a powerful in situ technique that contributes to our understanding of processes such as corrosion and corrosion protection, which are difficult to study using other methods [2]. The study concluded that NR is far from being used to its full potential in research. The current study is a step in that direction, applying operando NR to investigate how charging and discharging manifest inside LIB electrodes. This examination is difficult to perform with other techniques.

The elements in the 14$^{th}$ group of the periodic table, such as carbon, silicon, germanium, tin, and lead all have the same number of valence electrons. They also have the common property of reversibly storing lithium at low electrochemical potentials versus the Li metal reference potential. Therefore, they are ideal for use as active materials in negative electrodes of LIBs to store Li ions. Carbon in the form of graphite has long been used as an active material in negative electrodes of commercial LIBs because it retains its Li storage capacity even after prolonged cycling [28]. However, this high cycling stability comes at the cost of a relatively low lithium storage capacity [28]. Theoretically, graphite can store up to one lithium atom for every six carbon atoms ($LiC_6$) [28]. In practice, however, the storage capacity is lower [28]. Nowadays, this is insufficient, and materials with a higher storage capacity are required. Current research focuses on silicon and germanium, which can store Li up to $Li_{15}Si_4$ and $Li_{15}Ge_4$ [29-31], respectively. This is more than 22 Li atoms per 6 silicon (Si) or 6 germanium (Ge) host atoms. This is in contrast to graphite, which, as above mentioned, is still the ubiquitous active material for negative electrodes in commercial LIBs with a limited storage capacity of only 1 Li per 6 host carbon (C) atoms ($LiC_6$) [28,29]. Due to their high Li storage capacity, silicon and germanium are considered high-energy electrode materials. However, both active materials have not achieved a breakthrough in commercial LIBs due to capacity maintenance issues during cycling caused by volume changes. The ability to store large amounts of Li also presents challenges for Ge and Si electrodes, as they undergo significant volumetric changes during cycling. The mechanical stresses associated with these volume changes degrade the Ge and Si active material, pulverizing it during the process in the worst case [29]. These volume changes are crucial for LIB operation because they are related to the amount of stored energy as well as LIB integrity, performance, and safety. Therefore, knowledge of volume changes as a function of the state of charge (SOC) of the electrode is important to understand the overall behavior of energy storage materials.

The volume change of silicon during cycling was studied experimentally [3-5,8,10-12,22-35] and theoretically [36-39], as described in detail in Section 2. The behavior of germanium volume change during LIB cycling is nearly unknown. To address this knowledge gap, the current study presents the volume change of germanium during cycling as measured by operando NR during Li insertion and Li extraction.

In the case of experiments on silicon [3,4,8,10-12], it could not be clearly discerned whether the measured volume change stemmed solely from the cycled silicon material or contributions from battery-side reactions, such as the growth and shrinking of the SEI layer are included. In order to address this problem for the present studies on germanium, the current operando NR investigations use the ability of neutron scattering to discriminate isotopes to specifically determine the intrinsic volume change of the active material in the electrode as a function of SOC.

The active material under investigation in this study is prepared in the form of isotope multilayers (IMLs). An IML is a repetitive arrangement of bi-layers consisting of two different stable isotopes of an element with different neutron scattering lengths [18]. This approach is based on the idea that an IML arrangement transforms complex neutron reflectivity patterns into simpler ones due to the modulation of the volume of the active material by isotopes. This reduces the surface effects at the electrode-electrolyte interface in the NR. The IML arrangement produces a well resolved Bragg peak in the NR. The scattering vector position of this peak is mainly determined by the layer thickness and thus by the volume [18]. Volume changes inside the active electrode material during electrochemical cycling manifest as shifts in the scattering vector position of the Bragg reflex. As a model experiment, this study monitored the volume change of amorphous germanium film electrodes operando with NR during charging and discharging cycles. The active material of the working electrode consists of a Ge IML with 20 repetitions of a [$^{nat}$Ge/$^{73}$Ge] bilayer. $^{nat}$Ge and $^{73}$Ge denote germanium layers with a natural (7.8 % $^{73}$Ge) and an isotope enriched (95 % $^{73}$Ge) abundancy, respectively. We have previously used such IMLs to successfully determine in situ the Ge self-diffusion in Ge$_x$Si$_{1-x}$ materials [17-21]. This study extends the application of Ge IMLs to understand how Li changes the volume of active material during cycling. An advantage of the use of the two germanium isotopes as a proof-of-concept experiment is the large difference in nSLD leading to higher Bragg peaks in comparison to silicon ($^{nat}$Si/$^{29}$Si). Results on the identification of the lithiation mechanism based on this method are published in reference [15].

The work is organized as follows: Section 2 provides a review of the literature on reported and predicted volume changes during the cycling of silicon and germanium. Section 3 describes the materials, cells, devices, and measurement technique used. Section 4 presents and discusses the advantages and limitations of the isotope multilayer approach. Section 5 summarizes the findings.

## 2. Literature survey on volume changes during cycling of silicon and germanium electrodes

For silicon, which is valence isoelectronic with germanium and a semiconductor with similar properties, the literature reports theoretical [36-39] and experimental [3-5,8,10-12,32-35] investigations of volume changes during cycling. These studies employed atomic force microscopy (AFM) [32-34], operando synchrotron X-ray imaging [35], and operando NR [3-5,8,10-12]. Figure 1 illustrate these data. Note that the volume change can be expressed either as $V_{LixM}$ / $V_M$ or as $(V_{LixM} - V_M)$ / $V_M$, where M equals Si or Ge. The former expression is used throughout this work (see Figure 1).

When only the Li-Li and Si-Si interatomic distances of elemental lithium and silicon are considered, the following straightforward equation relating the relative volume change to the Li content in Li$_x$Si formed during lithiation can be obtained

$$\frac{V_{Li_xSi}}{V_{Si}} = \frac{x \cdot V_M^{Li} + V_M^{Si}}{V_M^{Si}} \qquad . \tag{1}$$

Here, $V_M^{Li}$ and $V_M^{Si}$ are the molar volumes of Li and Si, respectively, and x denotes the Li content expressed as x in Li$_x$Si. $V_{Si}$ is the molar volume of the silicon film without Li (x = 0). For germanium cycling, $V_{LixSi}$ and $V_{Si}$ must be replaced with $V_{LixGe}$ and $V_{Ge}$, respectively. The volume changes are plotted in Figure 1 with dashed black lines for silicon (Figure 1a) and germanium (Figure 1b) cycling.

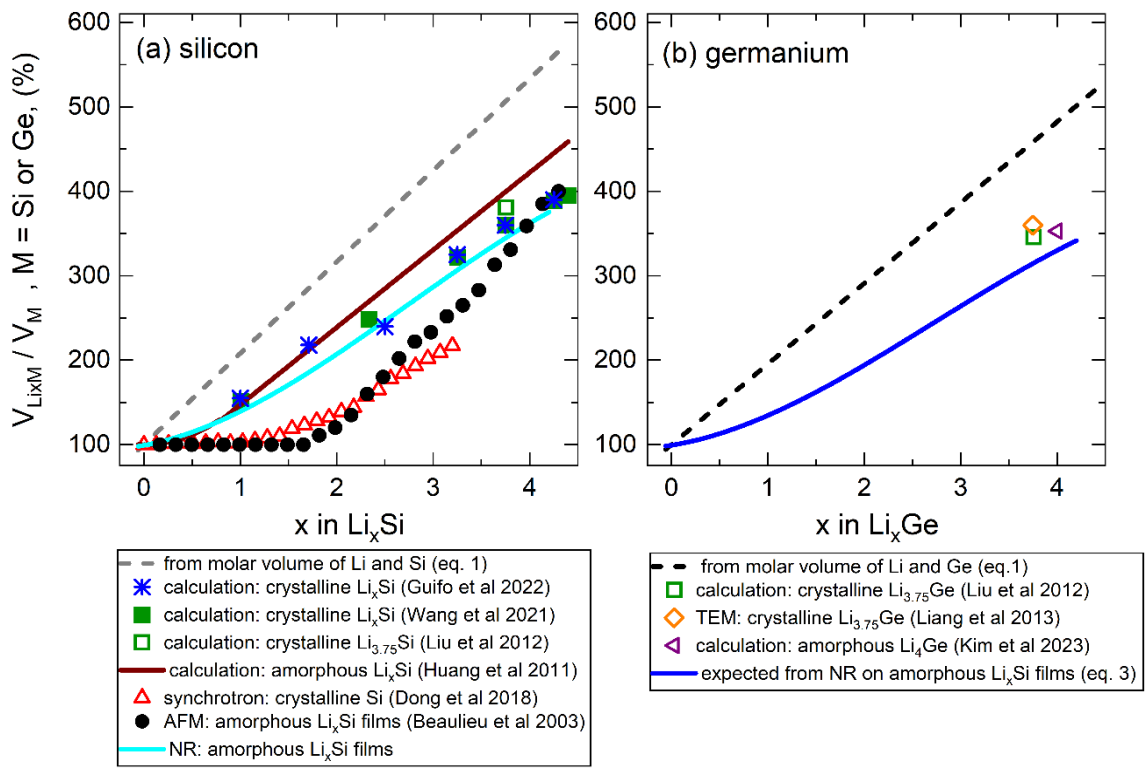

**Figure 1.** Volume changes during electrochemical electrode lithiation of **(a)** silicon ($V_{LixSi}/V_{Si}$) and **(b)** germanium ($V_{LixGe}/V_{Ge}$). See the main text for details.

First-principles calculations by Wang et al. [37] predict a relative volume expansion as plotted in Figure 1a with green squares. These data were obtained for the crystalline phases LiSi (x = 1), $Li_7Si_3$ (x = 2.33), $Li_{13}Si_4$ (x = 3.25), $Li_{15}Si_4$ (x = 3.75), $Li_{17}Si_4$ (x = 4.25), and $Li_{22}Si_5$ (x = 4.5) [37]. As can be seen, this results in a lower volume change rate. The formation of Li-Si bonds shrinks the volume. The same result is obtained by the calculations of Liu et al. for $Li_{15}Si_4$ [38] (open green squares), and by the newest first principle calculations of Guifo et al [39] (blue stars) for stable $Li_xSi$ phases. The molar volume of the stable and metastable crystalline phases of $Li_xSi$ is lower than that of elemental silicon and lithium [39]. A similar prediction is obtained for the amorphous $Li_xSi$ atomic network. Huang and Zhu [36] calculated the structure and energetics of amorphous $Li_xSi$ using molecular orbital theory (brown line). The values are also well below the black dashed curve.

Imaging [32-35] and reflectometry [3-5,8,10-12] were applied to determine the volume change in silicon electrodes experimentally during LIB cycling. The volume change in crystalline silicon was investigated by Breitung et al. [34]. They used in situ AFM to measure the change in electrode height while cycling crystalline silicon particles with a diameter of 100 nm embedded in binders and additives. The total thickness of the nano-Si electrode (including the current collector) was 34 μm. It was not possible to estimate the x-values (or SOC) from Breitung et al.'s report [34], so their values are not plotted in Figure 1a. Dong et al. [35] used operando synchrotron X-ray imaging to measure the volume change of micrometer-sized crystalline silicon particles mixed with binders and conductive additives. The silicon particles were mixed with carbon black and PVDF at a mass ratio of 45:45:10. Their data for 25 micron crystalline silicon particles are presented in Figure 1a with open red triangles. The respective x-values used in Figure 1a were obtained from their report indicating that the end of lithiation corresponds to x = 3.2. These results are well below the theoretical predictions (Figure 1a). Battery side reactions and the unknown Li content of the silicon particles may be responsible for the lower measured volume expansion.

AFM was also applied to experimentally investigate the volume change in amorphous silicon films [32,33]. Beaulieu et al. [32] used patterned and continuous amorphous silicon films approximately 350 nm thick, whereas Becker et al. [33] investigated 100 nm thick patterned amorphous silicon films. Beaulieu et al. found that the amorphous phase undergoes reversible shape and volume changes. Crystalline materials do not react in this way [32]. This difference was attributed to the homogeneous expansion and contraction that occurs in amorphous materials. Due to adhesion of the film to the substrate and the linear expansion of the film, the following relationship can be considered

$$\frac{V_{Li_xSi}}{V_{Si}} = \frac{d_{Li_xSi}}{d_{Si}} \qquad (2)$$

where $d_{Li_xSi}$ and $d_{Si}$ (x = 0) are the respective thicknesses of the $Li_xSi$ and Si films (x = 0) during cycling. The microscopy investigations performed on patterned amorphous silicon films [28] provided experimental validation of equation (2). The present analysis implicitly assumes isotropic expansion in the direction normal to the substrate. However, constraints imposed by the substrate may induce anisotropic deformation which is beyond the scope of this study.

Figure 1a plots the volume change of amorphous silicon measured with AFM for the first lithiation with black dots. The silicon was lithiated at constant current with a C rate of approximately 0.1 C [32]. The x-values for this data were calculated in this study by considering the cut-off potential of 0.0 V at the end of the first lithiation period to correspond to x = 4.4, as suggested in reference [32]. However, this assumption is questionable because not all of the charge delivered to the silicon electrode during the first lithiation reacts with silicon to incorporate Li. Some of the delivered charge may have been consumed in battery side reactions, such as the growth of the SEI layer and others. This is evidenced by the lack of volume change during the lithiation for x < 1.6 (Figure 1a), which Beaulieu et al. attributed to a side reaction involving trace amounts of water in the electrolyte. It was not possible to estimate the x-values (SOC) from Becker et al.'s report [33].

Finally, operando NR measurements were performed on amorphous silicon films. In this case, the neutron beams only scatter at two interfaces: the interface between the film and the substrate and the interface between the film and the liquid electrolyte. The interference of these reflected neutrons produces modifications in the NR pattern known as fringes. These fringes are the small oscillations that will be shown and discussed in Section 4.1. The interference fringes are the features of interest recorded in NR investigations of single films. The volume expansion of amorphous silicon films investigated by operando NR were reported in [3,4,8,10-12]. Our group carried out detailed operando measurements on the volume expansion of 70-nm-thick single, not isotope modulated amorphous silicon films during lithiation at current densities between 2 and 8 µAcm$^{-2}$. These results agree with those of ref. [4,8,10,11]. Figure 1a shows these averaged relative volume modifications as a function of SOC as a light blue line. These values agree well with those calculated ab initio (green filled squares and blue stars). The slight deviation to a lower volume may arise due to the following reasons: (i) not all charge stored is from Li$^+$ ions, (ii) some Li$^+$ is used to form a SEI layer on the silicon film surface, and not all Li is incorporated into silicon, and (iii) Li may fill pre-existing free volume. Note that the volume change can be determined from NR measurements independently of the exact lithiation model as discussed in [11].

In the case of germanium (Figure 1b), there are only a few published results. The dashed line shows the predicted volume change if the Li-Ge bonds are neglected (equation (1)). Liu et al. [38] calculated the volume of crystalline Li$_{15}$Ge$_4$ (x = 3.75) to be lower than that predicted by equation (1). First-principles density functional theory (DFT) calculations predict a volume expansion of 353% for amorphous Li$_4$Ge (x = 4) [40] which is plotted in Figure 1b with a brown square. Further, Liang et al. [41] report a volume change of V$_{Li_xGe}$/V$_{Ge}$ of 360% for a fully lithiated Ge nanoparticle with an initial diameter of 160 nm experimentally determined by TEM, plotted in Figure 1b as an orange rhombus. The initial Ge nanoparticle was crystalline, but lithiation leads to amorphization [41,42]. Full lithiation crystallizes amorphous Li$_x$Ge into the end phase of crystalline Li$_{15}$Ge$_4$ (x = 3.75), which is discussed further in Section 4. All these data for fully lithiation are in relatively good agreement.

The blue solid line in Figure 1b shows the expected volume change of germanium, considering the volume change in amorphous silicon, as determined experimentally from operando NR experiments on single films (light blue curve in Figure 1a). The results on silicon are transformed considering the slightly larger molar volume of germanium, using the following equation

$$\frac{V_{Li_xGe}}{V_{Ge}} = 1 + \left(\frac{V_{Li_xSi}}{V_{Si}} - 1\right) \cdot \frac{V_{cSi}}{V_{cGe}} = 1 + (0.2112 \cdot x + 0.2172 \cdot x^2 - 0.0264 \cdot x^3) \cdot 0.882 =$$
$$1 + 0.1862 \cdot x + 0.1916 \cdot x^2 - 0.0233 \cdot x^3 \qquad (3)$$

where V$_{cSi}$ = 12 cm$^3$mol$^{-1}$ and V$_{cGe}$ = 13.6 cm$^3$mol$^{-1}$ are the molar volumes of crystalline silicon and germanium, respectively.

In this context, it should be noted that the techniques used to measure volume changes in films during cycling may not accurately determine volume changes induced by lithium incorporation alone. Side reactions (e.g. the SEI layer, which is separate from the lithiated film electrode) may interfere. The volume change measured by AFM may be the sum of the volume change in silicon and the volume of the growing or shrinking SEI layer. Furthermore, extracting the Li$_x$Si volume change from classical NR measurements on a single silicon film (not isotope modulated) is complicated by the interference of neutrons scattered at the SEI and the film itself. The SEI layer affects the interface fringes between surface and substrate interface. Additionally, the roughening of the electrode surface that may occur during cycling will affect the interface between the SEI layer and the active material. This could reduce the amplitude of the fringes if the active material is a single layer. However, roughening of the interface between the SEI layer and the IML only marginally affects the Bragg peak because the single layers of the IML are buried. This work introduces a new approach that uses IML and operando NR to measure volume changes solely in the active material, independently of the presence or change of the SEI, and surface roughening.

## 3. Materials and Methods

The volume change of an amorphous germanium film electrode (Figure 2a) was monitored in operando with NR during the electrochemical incorporation and extraction of Li (lithiation and delithiation, respectively). The germanium film shows a Ge isotope modulation (Figure 2b). The active material of the working electrode consists of Ge IMLs with a 20-fold repetition of a [$^{nat}$Ge/$^{73}$Ge] bilayer. Figure 2c shows an image of the pristine electrode. Two electrodes with the following material sequence were scheduled: SiO₂ disk (3.2 mm) / 15 nm Cr / 159 nm Cu / {20 × [$^{nat}$Ge (7.0 nm)/$^{73}$Ge (7.0 nm)]} and SiO₂ disk (3.2 mm) / 15 nm Cr / 153 nm Cu / {20 × [$^{nat}$Ge (6.4 nm)/$^{73}$Ge (6.4 nm)]}. The Ge IMLs were produced by ion-beam sputtering on Cu current-collector films. The Cu and Cr films were deposited by magnetron sputtering on polished quartz (SiO₂) disks with a diameter of 1 inch. A 15 nm thick chromium adhesion layer was sputtered between the copper film and the quartz disk. These two electrodes give the same results as will be shown below, indicating a good reproducibility.

Ion-beam sputtering was performed using a commercial setup (IBC 681, Gatan, Irvine, California, USA). This setup is equipped with two Penning ion sources. $^{nat}$Ge layers were sputtered from a disc-shaped, polycrystalline germanium target (MaTecK, Germany) with a diameter of 2 cm, and $^{73}$Ge layers were deposited from a corresponding $^{73}$Ge enriched target (95% enrichment, MaTecK, Germany). Both targets can be installed in the setup simultaneously and used successively without breaking the vacuum. The Ge IML was deposited at room temperature with 5 keV and 180 µA of argon sputtering gas at an operating pressure of $8 \times 10^{-5}$ mbar. Before each deposition, the targets were pre-sputtered to remove possible atmospheric contaminants. After depositing the desired layer of the ML, deposition was interrupted by placing a shutter between the sputter target and the sample. The next target was then brought into the sputtering position and pre-sputtered for 30 seconds with the shutter closed. The base pressure, i.e., the pressure without the argon gas inlet to the Ar$^+$ ion cannons, was $2 \times 10^{-6}$ mbar. To minimize film contamination during sputter deposition, the IBC device was placed under argon overpressure in a custom-built glove box with water-cooled copper walls. The layer deposition rate was determined using X-ray reflectometry (XRR) measurements as described in reference [21]. The Cr and Cu films were deposited by DC magnetron sputtering at the Paul Scherrer Institute (PSI, Villigen, Switzerland) using argon sputter gas at room temperature. The base pressure was $2.7 \times 10^{-6}$ mbar, the operating pressure was $3.6 \times 10^{-3}$ mbar, and the sputtering power was 150 W.

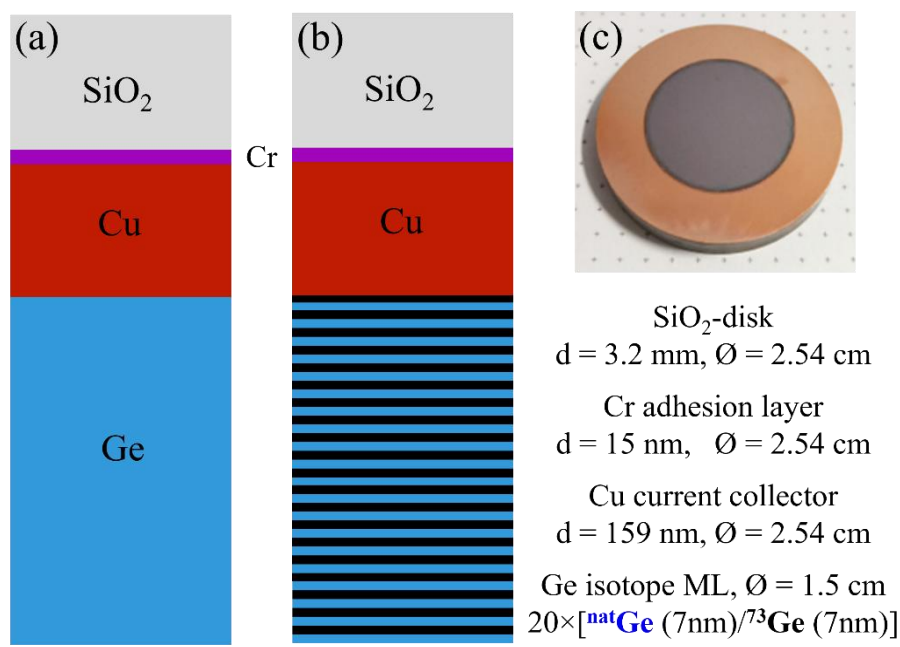

**Figure 2. (a,b)** Schematic cross-section of the Ge IML electrode. In **(a)**, the Ge isotope layers are not labeled. In **(b)**, the Ge isotope layers are labeled. The electrolyte is at the bottom and is in contact with a $^{nat}$Ge layer. **(c)** Photograph of the one-inch disk electrode. The active material (germanium) is the gray material in the middle of the disk. Information on dimensions of the different materials is given.

Electrochemical measurements were performed using a three-electrode electrochemical cell, as shown in Figure 3. The electrolyte was propylene carbonate (PC, Sigma-Aldrich, Taufkirchen, Germany, anhydrous, 99.7%) with 1 M LiClO$_4$ (Sigma-Aldrich, Taufkirchen, Germany, battery grade). Due to the large distance of 20 mm between the electrodes, no separator was used. The cell was assembled and disassembled in a glove box filled with argon gas. The content of O$_2$ and H$_2$O were less than 1 ppm. After disassembling the cell, the diameter of the lithiated zone on the electrode surface was visually determined to be about 15 mm, which coincides with the diameter of the cell tube filled with liquid electrolyte and the Ge IML. The counter and reference electrodes were lithium plates (1.5 mm thick, 99.9%, Alfa Aesar, Kandel, Germany). All reported germanium potentials $E_{we}$ refer to the lithium metal reference electrode. In this article, we will use the terms delithiation (extraction of Li from germanium) and lithiation (incorporation of Li in germanium) instead of the more common terms charging and discharging. Each lithiation/delithiation cycle begins with lithiation of the germanium electrode. Electrochemical studies were performed using a Biologic SP150 potentiostat and EC-Lab software version V11.43 (Biologic, Seyssinet-Pariset, France).

The cell was cycled at room temperature in constant-current (CC) mode with current densities ranging from 4 to 200 µAcm$^{-2}$ within an electrode potential window of 0.02 to 3.2 V. The low current density of 4 µA cm$^{-2}$ resulted in a very low lithiation rate of ≈ 0.023 C, where C-rate is defined as the inverse of the lithiation time in h$^{-1}$ units. This might be not of interest for common

LIB operation (e.g. in electric vehicles). Nevertheless, this low C-rate was investigated because it is closer to the stable thermodynamic behavior of the intrinsic volume change process.

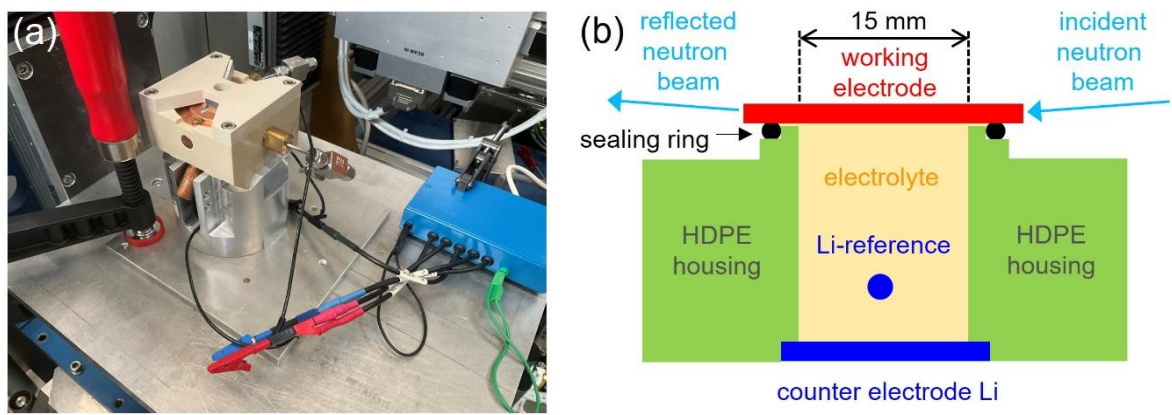

**Figure 3. (a)** Photograph of the electrochemical cell mounted on the AMOR reflectometer sample table. **(b)** Schematic cut through the electrochemical cell. The working electrode (marked in red) was a 1-inch quartz disk with the IML arrangement shown in Figure 2.

In general, reflectometry measures the specular reflected intensity of some type of radiation from a flat surface and deduces the related density profile as a function of vertical distance from the surface (depth profile). The reflecting power of an interface depends on the density contrast of the materials. These densities are often expressed as refraction indices. While optical light probes polarizability and X-rays probe electron density, neutrons used here interact with atomic nuclei. The nSLD is a measure of the average attraction or repulsion of a neutron by an ensemble of isotopes. This is important because isotopes of one element may have quite different interaction potentials, which are expressed as scattering lengths.

The "reflectivity curve or pattern" is the fraction of the intensity reflected as a function of the scattering vector, $q_z = (4\pi \sin\theta/\lambda)$, with the angle of incidence, $\theta$, and the neutron wavelength, $\lambda$. When the penetrating fraction of the beam encounters another interface, it is partially reflected and partially transmitted again. All partial waves leaving the surface interfere, so the resulting measurable intensity is a function of all interface contrasts, the distances between them, and $q_z$. In systems with many interfaces, complex, ambiguous and often difficult to interpret reflectivity patterns may evolve. In multilayers with a periodic arrangement of layers, the periodicity of the layer sequence produces a distinct Bragg peak in NR.

Reflectivity measurements are often analyzed by comparing a model based calculated reflectivity (simulation) to the measured reflectivity. The model essentially consists of a sequence of layers, each with a given nSLD and thickness (Figure 2). The relationship between nSLD and chemical composition must be based on external knowledge and other analytical techniques. However, for a variation of the isotope composition, variations in the nSLD profile can be related to a single parameter: the absolute density of the respective element. The current work exploits the controlled variation of the isotope composition together with the accurate determination of the layer thickness (and, consequently, the volume) of that isotope contrast. Specular NR probes the laterally averaged nSLD with a depth resolution down to atomic distances. This averaging means that no information can be deduced about lateral order on a small scale or even crystallinity.

As mentioned in the introduction, neutrons have a high penetration depth (several centimeters) in most materials and can easily penetrate the environment of the surface under investigation. This environment may be the chamber of an electrochemical cell [3-15] or a furnace [17-22]. Therefore, it is possible to perform these measurements on buried interfaces, even during operation, as long as the interfaces are sufficiently flat and large. Films on flat surfaces and functional devices based thereon are ideal candidates for this method.

NR simulations were performed using the Parratt32 and the Motofit software package, which are both based on the Parratt formalism [43,44]. The online calculator given in reference [45] was used to calculate the nSLDs. Both software packages, Parratt32 and Motofit, delivered identical results.

NR patterns were recorded in operando at the AMOR reflectometer (Apparatus for Multi-Option Reflectometry) and the Selene setup [22,24] located at the SINQ neutron spallation source (Schweizer Institut für Neutronenquellen) at the Paul Scherrer Institut (PSI) in Switzerland. The Selene focusing guide is used to achieve high-intensity specular reflectometry. Rather than using a highly collimated beam like conventional time-of-flight reflectometers, AMOR accepts a convergent beam covering a large angular range and the angular resolution is realized by the position-sensitive detector. This angle- and energy-dispersive setup allows for fast recording of specular NR signals. The raw data are stored as an event stream which allows for a flexible time-binning after the data collection. The time resolution used here is 1 min for high current densities and up to 20 min for lower current densities, respectively. The data reduction was performed as described in reference [24].

NR, XRR, grazing incidence XRD (GI-XRD), Raman spectroscopy, and Auger electron spectroscopy (AES) were performed at room temperature. XRR and GI-XRD were carried out on a Bruker D8 DISCOVER diffractometer (CuKα, 40 keV, 40 mA) (Bruker AXS Advanced X-ray Solutions GmbH, Karlsruhe, Germany) with an incident angle of 1°. Raman scattering was performed using a Bruker SENTERRA Raman microscope (Bruker, Rosenheim, Germany) with a 532 nm laser. For more information, see reference [46]. AES spectra were measured using a NanoSAM Lab microscope (Omicron GmbH, Taunusstein, Germany).

## 4. Results

### 4.1 Pristine Ge isotope multilayer electrode characterization

Various methods were used to characterize the pristine electrodes before NR investigations were done, in order to obtain information on the structural state. GI-XRD revealed that the copper current collector is polycrystalline. GI-XRD and Raman scattering showed that the Ge IMLs are amorphous, consistent with previous reports [17,21]. AES measurements demonstrated that the oxygen contamination of the germanium film is below the AES sensitivity limit of approximately 1 at.-%.

Figure 4 shows the NR pattern of the film electrode in an electrolyte-filled electrochemical cell in its pristine state (before the first lithium insertion). Small oscillations (fringes) are visible between a wave vector of 0.01 and 0.04 Å$^{-1}$. These fringes stem from neutron beam reflections at the two interfaces of the Cu current collector. The film thickness of the Cu current collector was determined using the Kiessig analysis, as described in references [47]. A Cu layer thickness of (153 ± 1) nm was obtained. A strong Bragg peak appears at a wave vector of $q_z = \approx 0.048$ Å due to Ge isotope modulation.

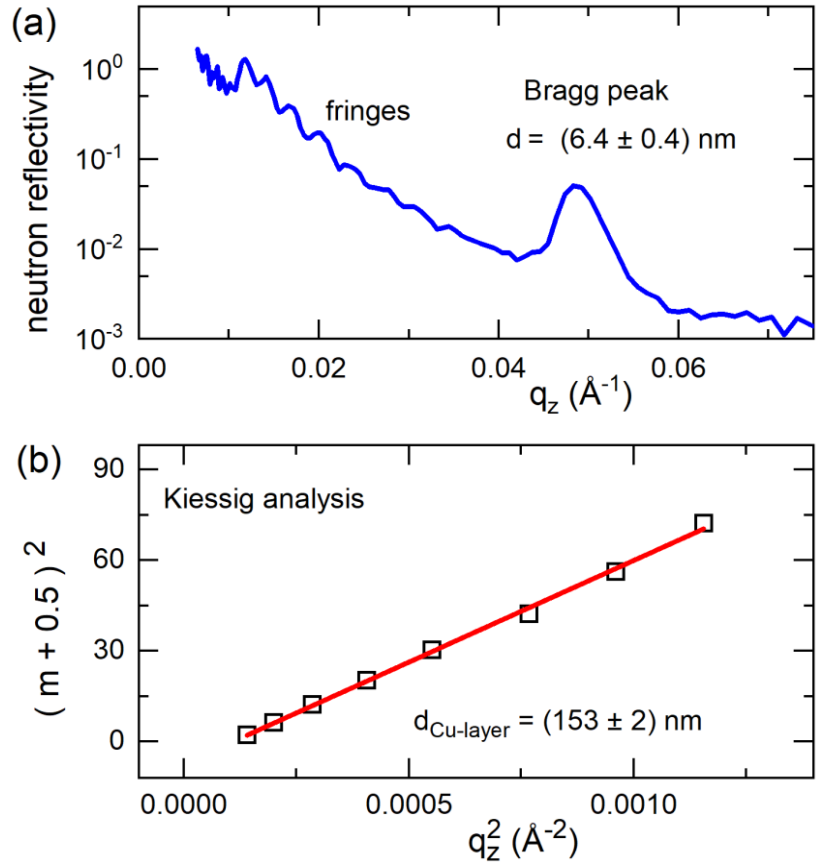

**Figure 4. (a)** NR pattern of the pristine Ge IML electrode that was mounted in an electrochemical cell filled with electrolyte. **(b)** Kiessig analysis of the fringes due to the Cu current collector; m is the fringe order.

Thorough NR simulations were carried out for data analysis (see below) with input data of Figure 4. As an example, Figure 5a shows the calculated nSLD depth profile of the electrode within an electrochemical cell filled with electrolyte for illustration. Figure 5b displays the corresponding simulated NR pattern with a Bragg peak due to Ge isotope modulation and fringes due to reflection at the Cu current collector. The very small single Cr layer has no significant influence in the scattering vector range displayed. The zero point is at the interface between the quartz holder disk and the Cr adhesion layer. The Bragg peak is located at a scattering vector of $q_z \approx 0.048$ Å$^{-1}$ for a single Ge layer thickness of 6.4 nm in the IML. The Bragg peak has a significant higher intensity than the fringes (logarithmic scaling). NR experiments and simulations have shown that the Cu current collector is not modified during LIB cycling, because Li does not alloy with copper [48,49], at least not at room temperature [49,50]. The information on volume modification during cycling is extracted from the location of the Bragg peak in $q_z$ (Figures 4a and 5b) by a combination of measurement and simulation.

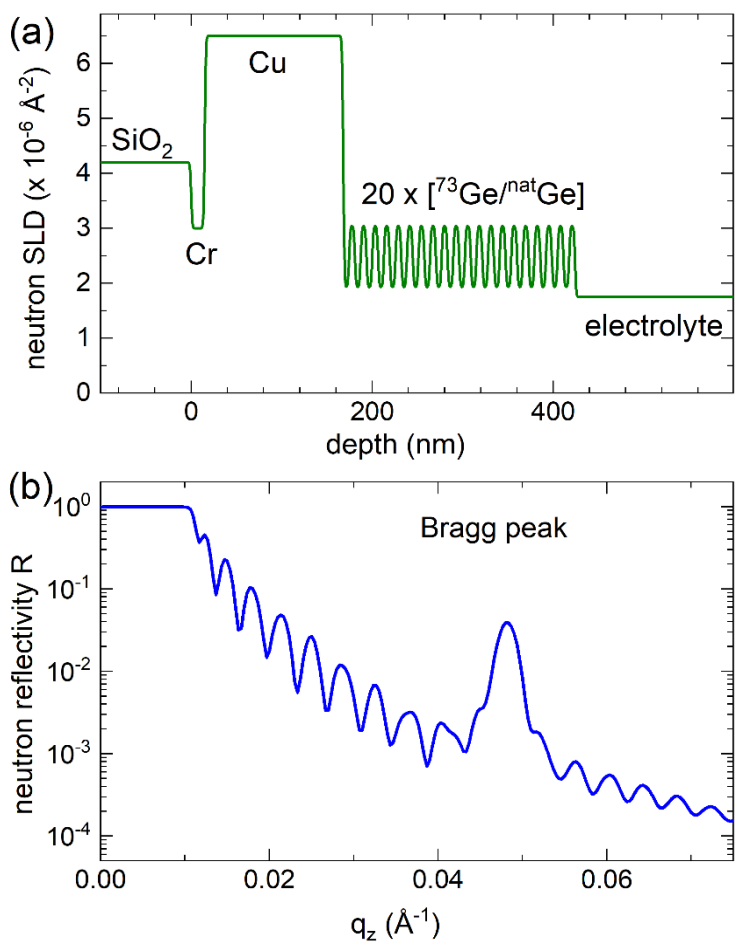

**Figure 5.** NR simulation of the pristine Ge IML electrode inside an electrochemical cell filled with electrolyte. **(a)** nSLD depth profile and **(b)** reflectivity of the electrode, R.

**4.2 Electrochemical lithiation and delithiation cycles during operando NR experiments**

Figure 6 shows the electrochemical measurements performed during the operando NR experiments on the 20×[$^{nat}$Ge(6.4 nm)/$^{73}$Ge(6.4 nm)] electrode. The lithiation and delithiation cycles were done at a CC density of 40 µA cm$^{-2}$ for the first 2.5 cycles. Lithiation was performed up to a low potential of 0.022 V, and delithiation was performed up to a potential of 3 V to ensure full germanium lithiation (100% SOC) and delithiation. The C-rate was approximately 0.22 C.

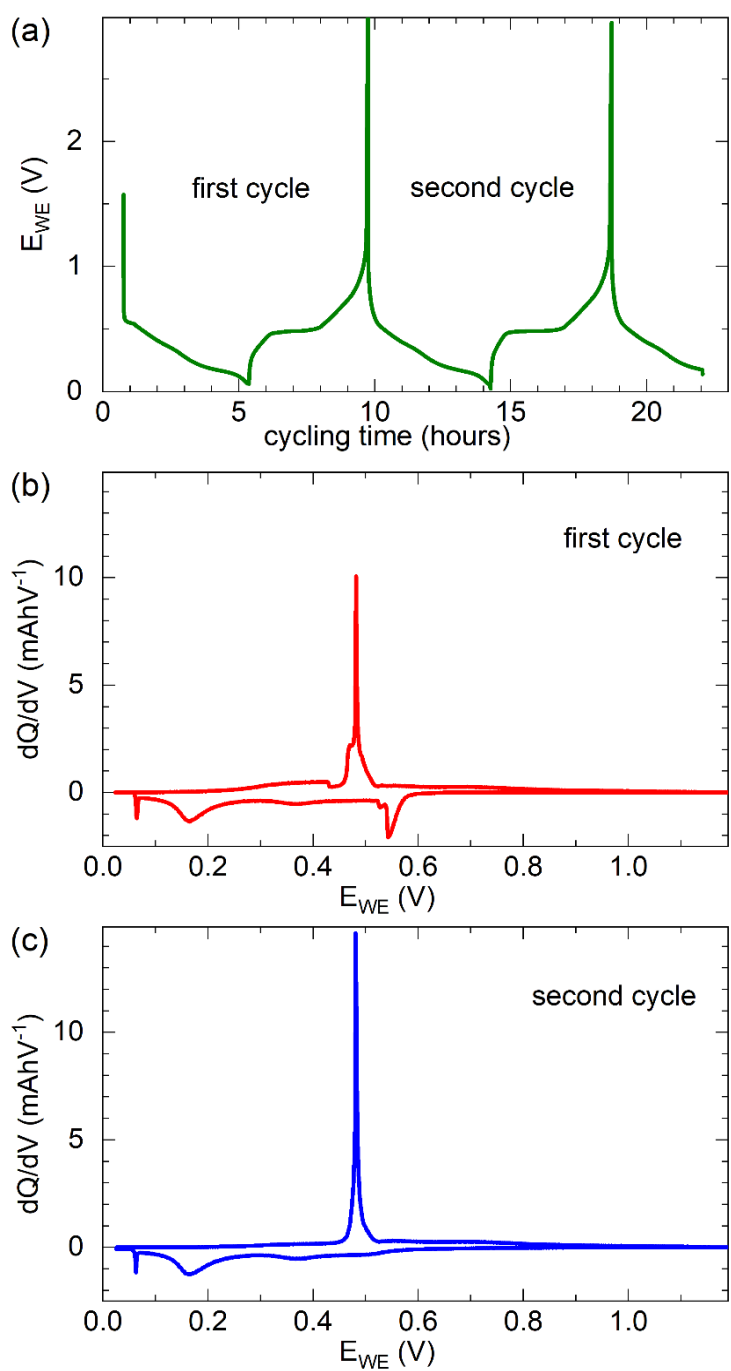

**Figure 6.** (a) Potential of the working electrode as a function of cycling time during the first 2.5 cycles with a CC density of 40 μA cm$^{-2}$ for the 20 × [$^{nat}$Ge (6.4 nm)/$^{73}$Ge (6.4 nm)] electrode. Corresponding dQ/dV curves as a function of potential obtained from the curves of (a) for the (b) first and (c) second cycle.

Figure 6a shows the potential of the working electrode as a function of cycling time, and Figure 6b and 6c plot the charge differential curves (dQ/dV) calculated from the curve in Figure 6a as a function of the potential. The potential profiles in Figure 6a are characterized by plateaus, indicating predominant Li$^+$ uptake and release potentials. These events are more clearly visualized as peaks during lithiation and delithiation in the dQ/dV plots, which represent the amount of charge (dQ) taken up or released within a small potential interval (dV). These plots generally resemble cyclic voltammetry curves. The lithiation (delithiation) process has negative (positive) values in the dQ/dV curves.

The first lithiation process (Figure 6b) shows three discernible dQ/dV (negative) peaks. Two of them are located at approximately 0.54 V, 0.17 V, and a small, sharp peak at approximately 0.06 V. The origin of the first two predominant Li$^+$ uptakes is currently unknown. The first (at 0.54 V) is sometimes attributed to the build-up of the SEI layer, similar to the case of amorphous silicon films [30,31]. Other authors believe that the SEI is predominantly formed at low potentials, e.g. at the Li$^+$ uptake peak at 0.17 V, because the electrolyte is more unstable at lower potentials [28-31,51]. However, there is agreement that the small, sharp peak at 0.06 V corresponds to the sudden crystallization of amorphous Li$_x$Ge when x approaches 3.75 (crystalline Li$_{15}$Ge$_4$ phase) as demonstrated by XRD [52-55] and by transmission electron microscopy (TEM) [41,42,52,55,56]. A similar crystallization process has also been observed for silicon lithiation [57-61] at a Li concentration slightly above x = 3.70. Thus, x = 3.75 can be thought to be the Li concentration for full Li$^+$ insertion into silicon and germanium [55].

The crystallization process also significantly influences the delithiation process. Delithiation is characterized by a sharp, strong, predominant Li$^+$ release peak at approximately 0.48 V, as obtained in this study (Figure 6b–c). During this release, a re-amorphization of the crystalline Li$_{15}$Ge$_4$ phase takes place. Thus, the occurrence of this peak in dQ/dV plots indicate that the Ge was fully lithiated to a Li content of x = 3.75. The crystallization process at 0.06 V and the amorphization process at 0.48 V also occur during the second cycle (Figure 6c).

The relative volume change was also determined for an electrode cycled without crystallization of Li$_x$Ge at x ≈ 3.75 (and consequently re-amorphization). To accomplish this, the CC cycling of the second 20×[$^{nat}$Ge(7.0 nm)/$^{73}$Ge(7.0 nm)] electrode was stopped at 0.18 V well before 0.06 V (Figure 7). As expected, the sharp, strong re-amorphization peak observed in Figure 6b–c at 0.48 V is absent in Figure 7b. The experiments shown in Figure 7a and 7b for the first cycle were performed with a lower constant current density of 4 µA/cm², delivering a C-rate of only 0.023 C. Figures 7c and 7d show the case for a higher current density of 153 µA/cm² that was selected for the second and third cycle, delivering a C-rate of 1.2 C. Crystallization and amorphization are absent in all cases.

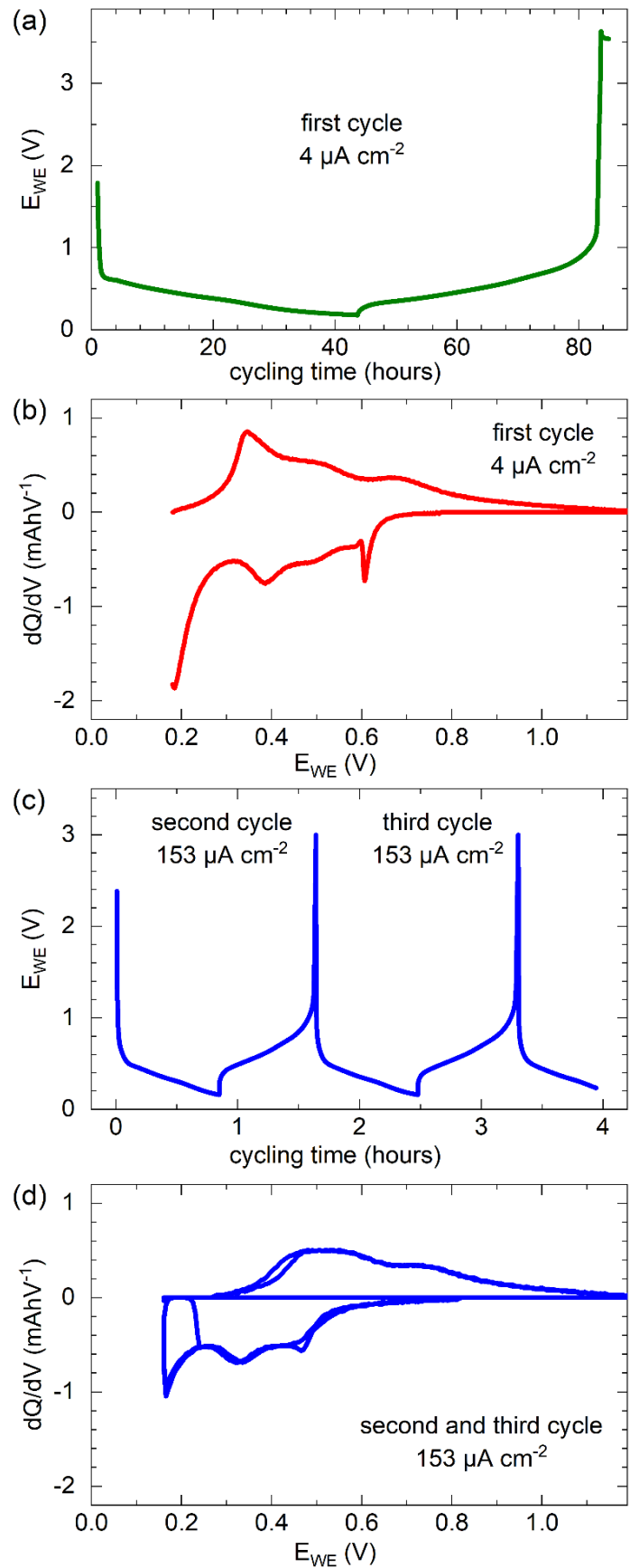

**Figure 7. (a)** Potential of the working electrode during the first cycle with a constant current density of 4 µA cm$^{-2}$ for the 20 × [$^{nat}$Ge (7.0 nm)/$^{73}$Ge (7.0 nm)] electrode. **(b)** Corresponding

dQ/dV curve as a function of potential obtained from the curve in **(a)**. **(c)** Potential of the working electrode during the second and third cycle with a constant current density of 153 µA cm$^{-2}$. **(d)** The dQ/dV curves obtained from the curves of **(c).**

Figure 8 shows the gravimetric capacity and the columbic efficiency during cycling. The electrode gravimetric capacity was determined using

$$C = \frac{Q}{m} = \frac{I \cdot t}{m} = \frac{I \cdot t}{\pi \cdot r^2 \cdot d \cdot \rho}, \tag{4}$$

where *I, t, r, d*, and *ρ* are the current, the time interval of the lithiation and delithiation processes, the radius of the lithiated electrode surface, the thickness, and the mass density (see below) of the whole Ge IML, respectively.

The gravimetric capacity of the 20 × [$^{nat}$Ge (6.4 nm)/$^{73}$Ge (6.4 nm)] electrode (Figure 8a) attains values close to those reported for Li$_{15}$Ge$_4$, which is thought to be the maximum practical capacity of germanium [55]. This points to the fact that Li is incorporated into and released from germanium during the NR experiments. As expected, the Li capacity of the 20 × [$^{nat}$Ge (7.0 nm)/$^{73}$Ge (7.0 nm)] electrode (Figure 8a) at a low current density of 4 µA cm$^{-2}$ is slightly lower due to the higher cut-off potential for lithiation used. At the higher current density of 153 µA/cm we see a reduction due to kinetic limitations. The coulomb efficiency in Figure 8b is always slightly below 100% for both electrodes, meaning a slightly smaller amount of charge is extracted than inserted. Minor battery side reactions consume a small portion of charge in each cycle (less than 10%). This means that the pristine state cannot be completely recovered by delithiation, a fact that seems to be ubiquitous for LIB electrodes. Despite this issue, in the current work it is assumed that the charge delivered to or extracted from the electrode is approximately equal to the Li content in germanium, represented by x in Li$_x$Ge. This assumption is critically discussed in Section 4.4.

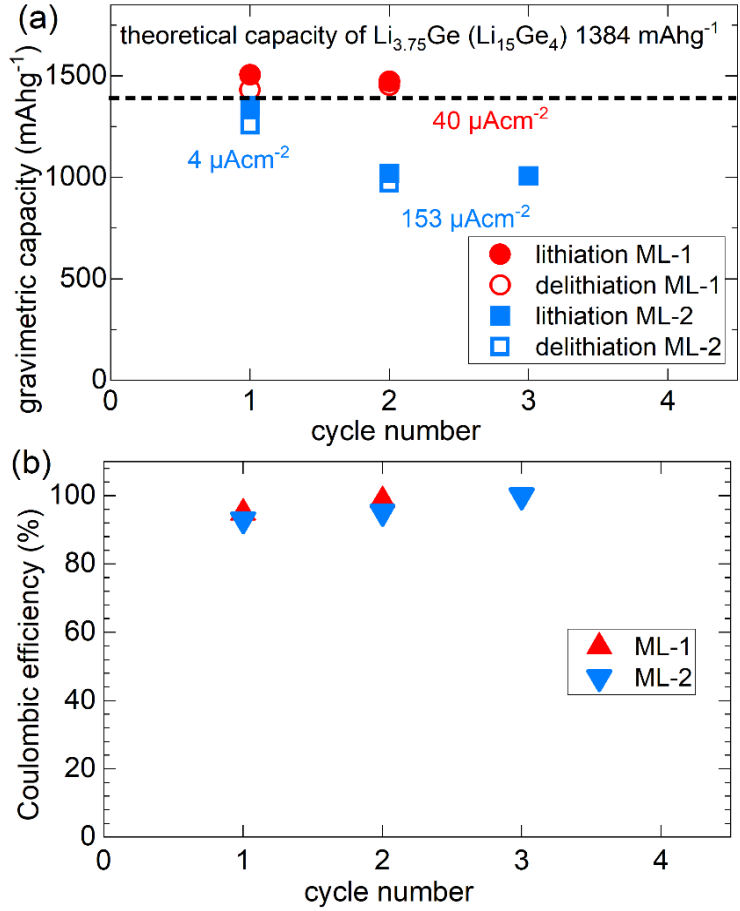

**Figure 8. (a)** Gravimetric capacity of Li$^+$ insertion and release of Ge IMLs as a function of cycle number. ML-1 denotes to the 20 × [$^{nat}$Ge (6.4 nm)/$^{73}$Ge (6.4 nm)] electrode (Figure 6) and ML-2 to the 20 × [$^{nat}$Ge (7.0 nm)/$^{73}$Ge (7.0 nm)] electrode (Figure 7). **(b)** The coulombic efficiency for the data in **(a)**.

The amount of Li, x, inserted into or extracted from the Li$_x$Ge IML electrode during cycling was calculated using the following equation:

$$x = \frac{N_{Li}}{N_{Ge}} = \frac{\frac{Q}{e}}{\frac{m_{Ge}^{film}}{m_{Ge}^{atomic-mass}}} = \frac{I \cdot t \cdot m_{Ge}^{atomic-mass}}{e \cdot m_{Ge}^{film}} = \frac{I \cdot t \cdot m_{Ge}^{atomic-mass}}{e \cdot \rho_{Ge}^{film} \cdot (d \cdot \pi \cdot r^2)} \qquad (5)$$

where N$_{Li}$ and N$_{Ge}$ are the total number of Li and Ge atoms in the electrode at a given SOC, Q is the amount of Li$^+$ charge in the film at that SOC, m$_{Ge}$$^{atomic-mass}$ ≈ 1.2 × 10$^{-22}$ g is the mass of a single Ge atom, m$_{Ge}$$^{film}$ is the mass of the Ge film (without Li), e = 1.6 × 10$^{-19}$ As is the elemental charge, d is the thickness of the entire Ge IML (without Li), and r is the radius of the Ge film. Finally, ρ$_{Ge}$$^{film}$ ≈ (4.5 ± 0.2) gcm$^{-3}$ is the mass density of the Ge film (without Li), as determined experimentally from NR measurements. To achieve this value, a 30 × ($^{nat}$Ge (7 nm)/$^{73}$Ge (7 nm)) ML was deposited directly on a 3 mm thick Si wafer (i.e. without Cr and Cu layer) and NR was performed on this layer in air. Here, the edge of total reflection appears,



which is used to determine the mass density by simulations to (4.5 ± 0.2) gcm$^{-3}$. This is within the range of reported mass densities of sputtered or evaporated amorphous Ge films, which are between 3.15 and 5.35 gcm$^{-3}$ [62,63]. The mass density of 4.5 ± 0.2 gcm$^{-3}$ obtained from NR for amorphous germanium films compared to the value of 5.35 gcm$^{-3}$ for crystalline germanium may indicate the presence of roughly 15 % free volume.

**4.3 Determination of relative volume change with operando NR measurements**

Figure 9 shows typical NR curves measured operando during the first lithiation of a 20×[$^{nat}$Ge(6.4nm)/$^{73}$Ge(6.4nm)] electrode at various SOCs. Different SOCs can be expressed as the amount of Li incorporated into the Ge IML (x in Li$_x$Ge electrode) as obtained from equation (5) and electrochemical measurements. A clear shift of the scattering vector position of the Bragg peak, q$_z$, to lower values with increasing x is visible. The modification in intensity is discussed below. Note that a small portion of the Bragg peak (about 12 %) is not shifted. This is likely due to portions of the electrode not being in contact with the electrolyte, as parts are covered by the cell's sealing components. However, this area is also detected by NR. Consequently, a fraction of the electrode area appears electrochemically inactive, as indicated by the presence of an unshifted Bragg peak contribution that persists during lithiation. Since NR averages laterally over the illuminated area, the measured signal represents a superposition of active and inactive regions. This may bias the extracted volume changes, as Li insertion into and extraction from the Ge IML are limited to the regions in direct contact with the electrode. Post-mortem visual inspection inside the glovebox, following cell disassembly, indicates that up to 10 % of the Ge IML does not participate in electrochemical cycling. This estimate is consistent with the ratio of the lithiated to pristine intensity of the Bragg peak shown in Figure 9b. Consequently, the Li content in the Ge IML, as derived from the measured lithiation and delithiation currents, is probably overestimated by approximately 12 %. Section 4.4 provides a more detailed discussion of this aspect.



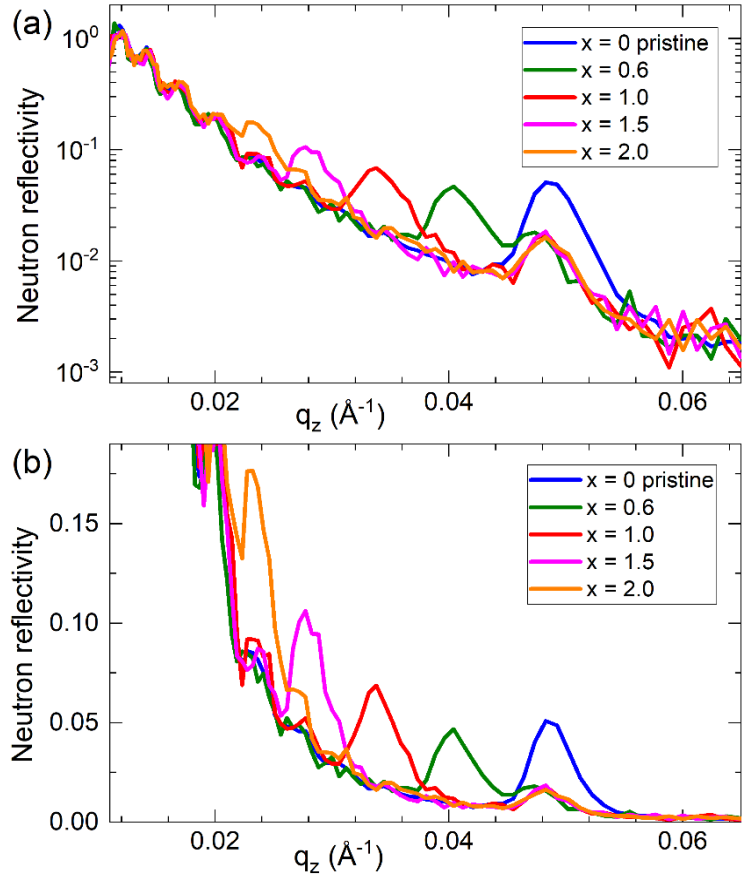

**Figure 9.** NR patterns of the 20×[$^{nat}$Ge(6.4nm)/$^{73}$Ge(6.4nm)] electrode for different x in Li$_x$Ge measured operando during first lithiation with a constant current density of 40 µAcm$^{-2}$. **(a)** logarithmic scaling and **(b)** linear scaling.

Figure 10 shows contour plots of the NR intensity as a function of scattering vector and SOC during the first cycle of the Ge IML as obtained from operando NR measurements (Figure 10a,c) and NR simulations (Figures 10b,d). The SOC is related to the Li content in the amorphous Ge IML, the cycling time, and the Ge IML electrode potential, as explained in the SI of reference [15]. The NR intensity is shown on color scale. The vertical bars in Figures 10a and 10b stem from the NR fringes originating from the copper current collector film. To more clearly highlight the changes in the NR patterns associated with the Bragg peak, Figures 10b and 10d present data where the NR curve of the electrode without the Ge IML (i.e., the curve lacking the Bragg peak) has been subtracted from each corresponding NR curve. In both of these plots some vertical bars are still visible. These are due to Li incorporation and removal into/from the Ge IML, which slightly also influence the fringe position. The copper current collector does not participate in cycling at electrode potentials below 3.5 V used in this study.

Figure 10 shows good agreement between the simulated and measured data. This agreement is also evident in Figure 11, which plots characteristics of the Bragg peak as a function of the SOC



during the first lithiation. The measured and theoretically predicted behavior of the Bragg peak amplitude (Figure 11a), Bragg peak baseline (Figure 11b), and Bragg peak full width at half maximum (FWHM) (Figure 11c) are similar. The increase in Bragg peak amplitude and baseline, along with the decrease in FWHM with increasing SOC, can be attributed to the shift of the Bragg peak toward lower $q_z$ values.

Plots as shown in Figure 10 are useful for identifying the lithiation and delithiation mechanism of active electrode materials, as shown in reference for germanium [15]. There, a homogeneous mechanism was identified, as reflected in the continuous shift in the scattering vector position of the Bragg peak as a function of SOC. Such a mechanism is a perquisite for the extraction of the correct volume modification by Bragg peak shift. Here, the term "homogeneous" refers to the evolution of the Li concentration as a function of depth within the IML film, rather than to any lateral Li distribution. Since NR is a laterally averaging technique, it does not resolve in-plane variations. During homogeneous lithiation, Li ions entering the film rapidly diffuse throughout the entire film thickness, resulting in an approximately uniform Li concentration that increases with SOC. For more details see ref. [15].



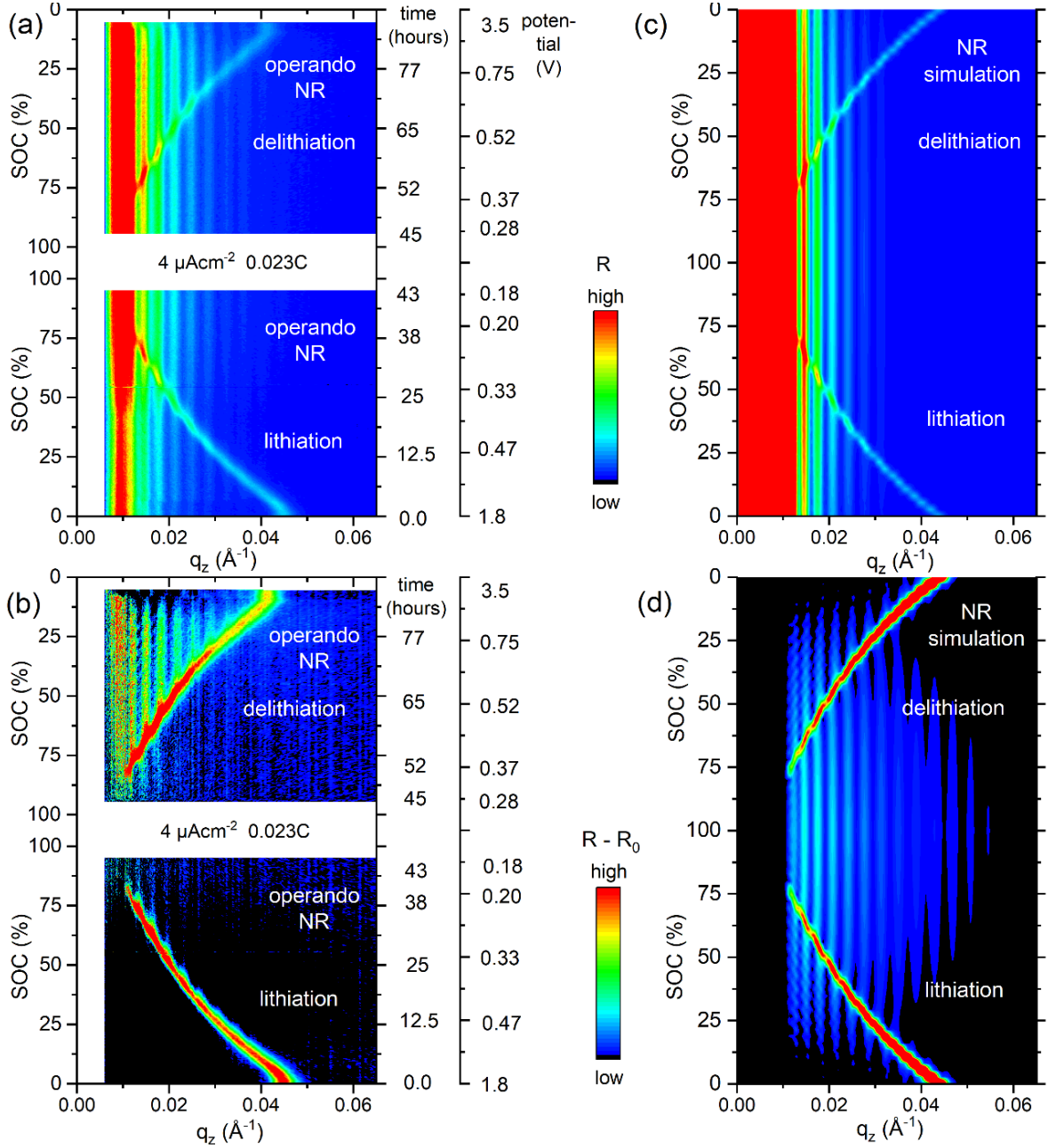

**Figure 10.** Contour plots of NR patterns as a function of scattering vector and of SOC during the first cycle for (**a,b**) operando measurements of the 20×[$^{nat}$Ge(7.0nm)/$^{73}$Ge(7.0nm)] electrode with a constant current density of 4 µAcm$^{-2}$, and for (**b,d**) corresponding simulations. Cycling times and electrode potential are indicated. The NR intensity is displayed on color scale. High to low intensity is displayed from red, over orange, yellow, blue to black color. (**a,c**) R and (**b,d**) R-R$_0$ representation, where R$_0$ denotes the NR intensity of the pristine state (SOC = 0%) in the absence of the Bragg peak (for details see the SI of reference [15]).



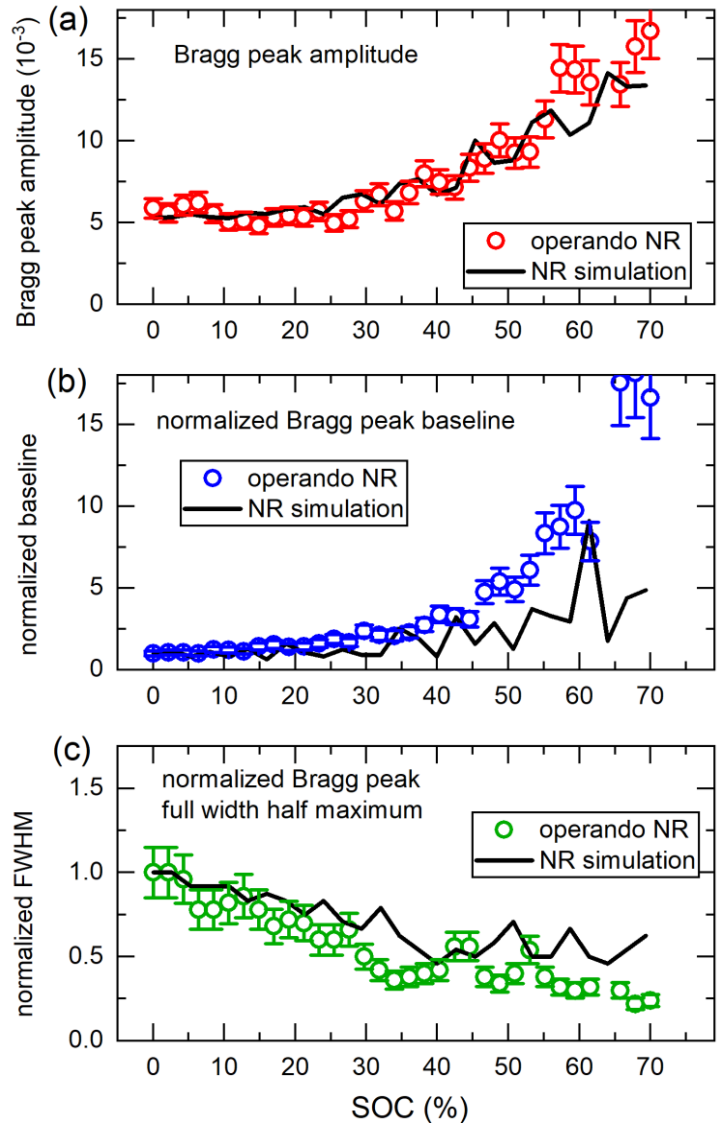

**Figure 11**. The evolution of relevant characteristics describing the NR Bragg peak as a function of SOC as obtained from measured and simulated NR patterns. Shown are **(a)** the amplitude of the Bragg peak, **(b)** the normalized baseline and **(c)** the FWHM. The data are extracted from Figure 10 b, d.

Figure 12a plots the scattering vector position of the Bragg peak against the Li content x in $Li_xGe$ of the 20×[$^{nat}$Ge(7.0nm)/$^{73}$Ge(7.0nm)] IML as obtained from operando NR measurements during the first lithiation at a constant current density of 4 µAcm$^{-2}$. The Li content x was calculated using equation (5). A continuous shift of the Bragg peak location in $q_z$ to lower values is observed with increasing Li content x as indicated in Figure 9 and Figure 10. This reflects an increase in volume if a homogeneous lithiation mechanism is assumed, as proven to be present in germanium in reference [15]. We can detect the Bragg peak necessary for analysis from NR patterns down to a scattering vector of 0.012 Å$^{-1}$ (x = 3). For lower values the background intensity is too high for a proper analysis (Figure 9).



As already mentioned, NR experimental results and further simulations revealed that the nSLD of the substrate remains unchanged during cycling, while the nSLD of the $Li_xGe$ layers undergoes significant modifications in addition to the film thickness by Li incorporation. In this context, the $q_z$ position of the Bragg peak is mainly determined by the $^{nat}Ge/^{73}Ge$ IML thickness and hence by the volume. Consequently, we assume for analysis in a first approach that the position of the Bragg peak in $q_z$ is approximately independent of the nSLD of each of the two Ge layers $^{nat}Ge$ and $^{73}Ge$. This would allow a straightforward and easy analysis of the results. Figure 12b shows the relative volume change $V_{LixGe}/V_{Ge}$ as a function of $q_z$ position as calculated by Parrat32 (orange curve). As an approximation, the nSLD values of the two single layers ($Li_x{}^{nat}Ge$ and $Li_x{}^{73}Ge$) are kept constant during lithiation at the initial (pristine) value of $3.04 \times 10^{-6}$ Å$^{-2}$ for $^{nat}Ge$ and $1.02 \times 10^{-6}$ Å$^{-2}$ for $^{73}Ge$, respectively. The thickness of the two single layers is simultaneously increased as an input parameter of the simulation, giving the corresponding position of the Bragg peak in $q_z$. The relative volume change is calculated from the change in film thickness according to equation (2). Combining now the simulation of Figure 12b with the experimental data of Figure 12a we obtain the desired relative volume change as a function of x in Figure 12c (orange curve). The volume expansion is identical to the assessed result from equation (3) for x < 1, but shows significant deviation to higher values.

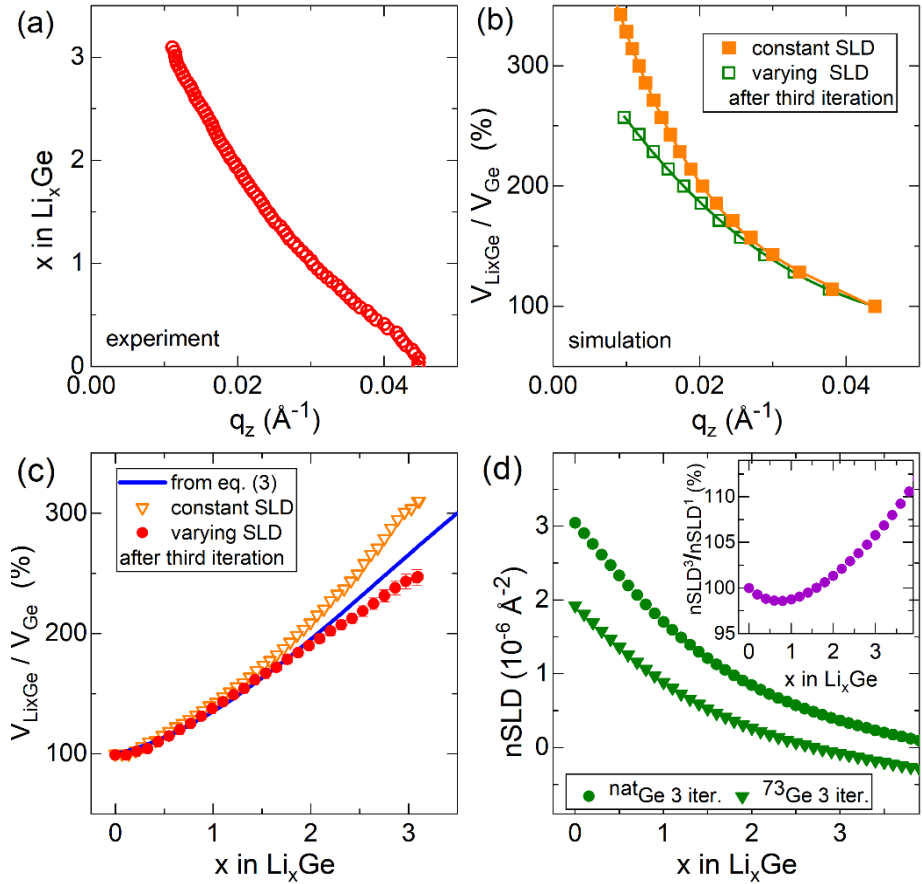

**Figure 12.** (a) Li content x in $Li_xGe$ as a function of the scattering vector position $q_z$ of the Bragg peak as determined experimentally during the first lithiation. (b) Relative volume change as a function of the scattering vector position $q_z$ of the Bragg peak as obtained from simulations.



**(c)** Relative volume change as a function of Li content x. **(d)** nSLD as a function of Li content x. The inset in **(d)** shows the ratio of the nSLD of the third and first iteration to illustrate the difference. For more details it is referred to the text.

For an improvement of the results, we consider now the influence of the nSLD on the position of the Bragg peak. A lower nSLD will shift the Bragg peak to slightly lower $q_z$ values. The nSLD of the $Li_x{}^{nat}Ge$ layers during cycling is calculated to

$$nSLD_{Li_xGe} = \frac{b_{Ge} + x \cdot b_{Li}}{V_{Li_xGe}} \tag{6}$$

where $b_{Ge} = 8.18 \times 10^{-15}$ m is the average neutron scattering length for all germanium isotopes in the natural germanium layer, $b_{Li} = -1.9 \times 10^{-15}$ m is the neutron scattering length for all Li isotopes, and $V_{LixGe}$ is the volume of the $Li_xGe$ chemical formula unit.

The nSLD of the $Li_x{}^{73}Ge$ layers is calculated accordingly to

$$nSLD_{Li_x{}^{73}Ge} = \frac{(0.95 \cdot b_{73Ge} + 0.05 \cdot b_{Ge}) + x \cdot b_{Li}}{V_{Li_x{}^{73}Ge}} \tag{7}$$

where $b_{73Ge} = 5.02 \times 10^{-15}$ m is the neutron scattering length for $^{73}Ge$ isotopes. Since the $^{73}Ge$ enrichment in the $^{73}Ge$ layers is only 95%, a correction factor is necessary. $V_{Lix73Ge}$ is the volume of the $Li_xGe$ chemical formula unit. In our case, $V_{Lix73Ge} = V_{LixGe}$.

With the use of equation (2), the volume change was calculated as follows

$$V_{Li_xGe} = V_{Ge} \cdot \left(\frac{V_{Li_xGe}}{V_{Ge}}\right) = \left(\frac{V_M^{Ge}}{N_A}\right) \cdot \left(\frac{d_{LixGe}}{d_{Ge}}\right) = \left(\frac{M_{Ge}}{\rho_{Ge} N_A}\right) \cdot \left(\frac{d_{LixGe}}{d_{Ge}}\right) \tag{8}$$

where $V_M^{Ge}$ is the molar volume of the amorphous germanium film, $N_A$ is the Avogadro constant, $M_{Ge} = 72.59$ g mol$^{-1}$ is the molar mass of Ge, and $\rho_{Ge} \approx (4.5 \pm 0.2)$ gcm$^{-3}$ is the mass density of the amorphous germanium film (without Li), as determined by NR.

The problem is now that the change of the nSLDs during cycling depends on the volume change according to equations (6) and (7). In order to calculate by Parratt32 the relative volume change as a function of $q_z$ position (Figure 12b), the correlation of volume and nSLD according to equations (6) and (7) has to be included. Consequently, the volume change has to be determined in an iterative way.

As a starting value for the first iteration step, we used the relative volume versus x dependence given by equation (3) (blue line in Figure 1b) to calculate the nSLD of the $Li_xGe$ layers as a function of Li content x by equations (6) and (7) (see Figure 12d). With these data, the Parratt32 simulation was done to determine the volume change as a function of the $q_z$ position and afterwards using Figure 12a the volume change as a function of x. This procedure was repeated two times. The result after the third iteration is shown in Figure 12c as red dots. As shown in Figure 12d, the nSLD values obtained for the three iterations differ only marginally. The inset of Figure 12d shows the difference between the third and first iteration for the $^{nat}Ge$ layer. The difference is smaller than -2% at low Li content, increasing to about +10% at high Li content. Consequently, further iterations are not necessary.

From the red curve in Figure 12c it is seen that at high $q_z$ values the volume change for high values of x is now lower than that of the other two curves. At the highest value of x = 3 the volume change is lower by about 20 % for the result of the iterative procedure than for the result with constant nSLD. This indicates that the iteration procedure is necessary to get exact results.

The complete results for the volume changes as a function of Li content x during cycling for higher cycles and for different charge densities / C-rates are plotted in Figure 13 and 14 for the two types of IML investigated.

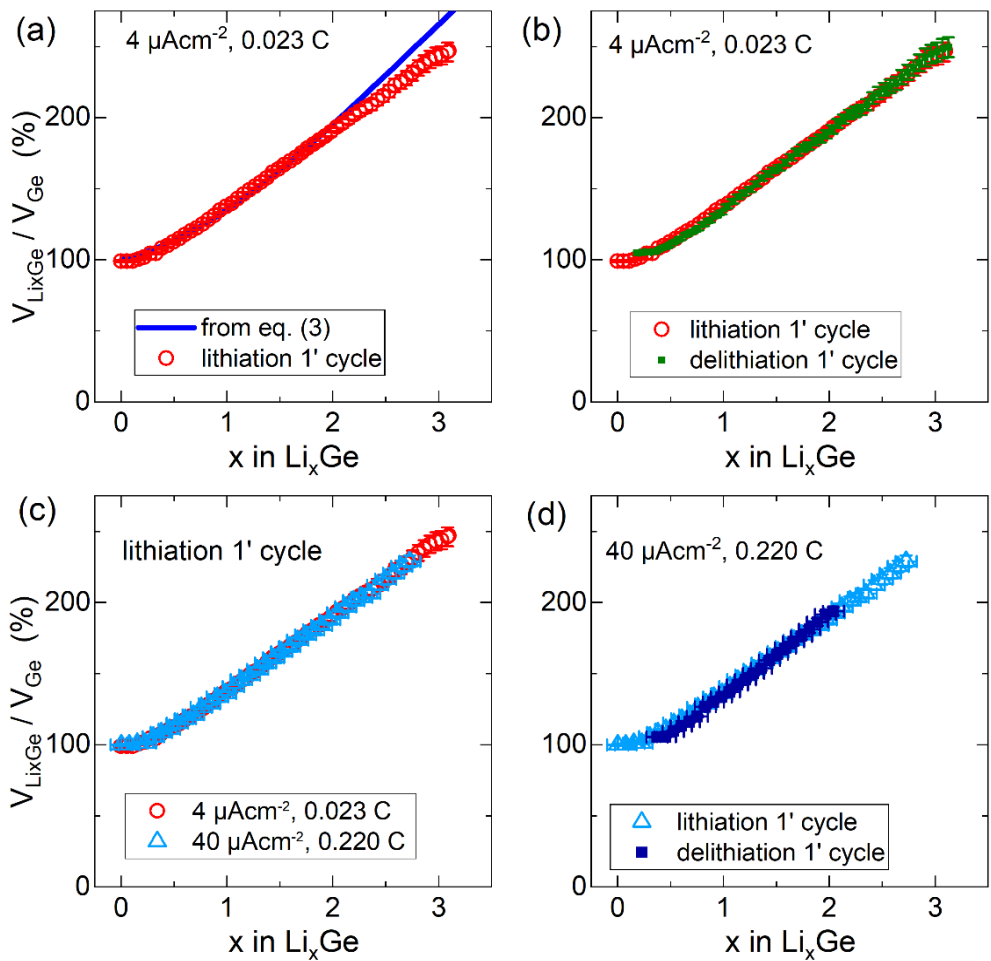

**Figure 13. (a)** Volume change of the amorphous Ge IML with $d_0$ = 7.0 nm at a C-rate of 0.023 C during first lithiation (red circles) compared to the predicted volume change of amorphous Ge from Si NR data (equation 3). **(b)** Volume change of the IML with $d_0$ = 7.0 nm during first lithiation (red circles) and delithiation (green squares) for the first cycle with a C-rate of 0.023 C. **(c)** Volume change during the first lithiation of the IML with $d_0$ = 7.0 nm with a C-rate of 0.023 C (red circles) and with $d_0$ = 6.4 nm with a C-rate of 0.22 C (blue triangles). **(d)** Volume change of the ML with $d_0$ = 6.4 nm during first lithiation (blue triangles) and delithiation (dark blue squares) at a C-rate of 0.22 C.

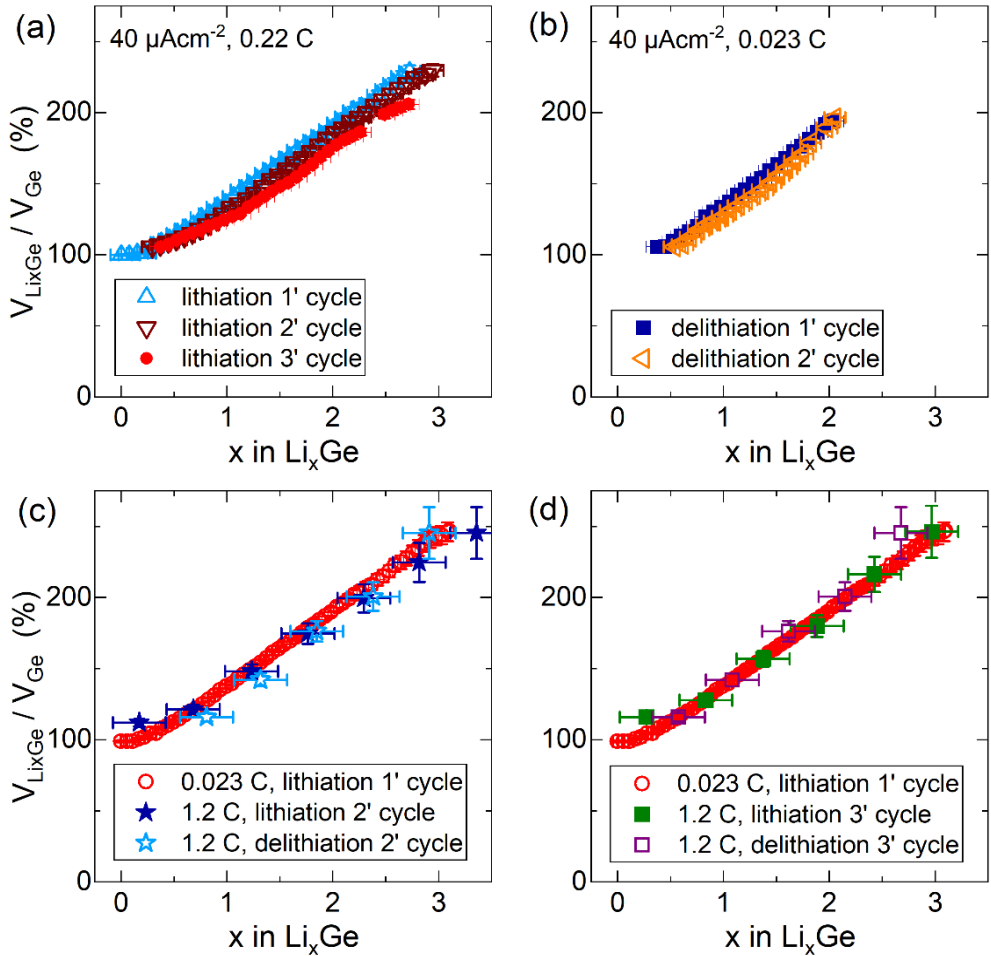

**Figure 14. (a)** Volume change of the IML with $d_0 = 6.4$ nm during lithiation with 0.22 C for the first (unfilled blue triangles), second (brown triangles), and third (red circles) cycle. **(b)** Volume change of the IML with $d_0 = 6.4$ nm during delithiation with 0.22 C for the first (filled dark blue squares) and the second cycle (orange triangles). **(c)** Volume change of the IML with $d_0 = 7.0$ nm during the second cycle with 1.2 C (dark blue stars for lithiation and unfilled blue stars for delithiation) compared to the first cycle with 0.022 C (unfilled red circles). **(d)** Volume change of the IML with $d_0 = 7.0$ nm during the third cycle with 1.2 C (green squares for lithiation and purple squares for delithiation) compared to the first cycle with 0.022 C (red circles).

Figure 13a shows the relative volume change during the first lithiation with the lowest current density applied (4 µA cm$^{-2}$, 0.023 C) for the Ge IML with 7 nm thick individual layers. The NR Bragg peak could be discerned up to a Li content of $x = 3$ (Li$_3$Ge), resulting in a volume change of 250%. Up to a Li content of $x \approx 2$, the volume change is identical to that derived from operando NR measurements on amorphous Si single films according to equation 3 (blue line in Figure 13a). For a higher Li content, the volume increase deviates to lower values. A possible explanation might be that in amorphous Ge more free volumes exist than in Si. The free volumes are filled by Li during lithiation. This will reduce the relative volume change at higher

Li content. Another reason might be side reactions, such as the growth and degrowth of the SEI layer, which is discussed later in section 4.4.

Figure 13b shows that the volume decreases during the first delithiation period (green squares) is identical to the volume increase during the first lithiation. Thus, amorphous germanium exhibits reversible swelling and shrinking during the first cycle. The same is true for the Ge IML with a shorter bilayer thickness and for faster cycling at higher charge densities (Figures 13c and 13d). As can be seen, delithiation during the first cycle does not restore the volume of the initial pristine state (Figures 13b and 13d). This can be attributed to irreversible charge trapping in the electrode [16].

As described in Section 4.2, the IML with a single layer thickness of 6.4 nm was cycled across a presumed crystallization and re-amorphization region (Figure 6). In this region there is no Bragg peak visible in the NR patterns (Figure 10a,c and Figure 10b,d). Consequently, the volume change could not be measured. The reason is unclear at the moment. However, the volume changes determined at other SOCs outside this region are comparable to those obtained when crystallization and amorphization were suppressed (Figure 13c). This suggests that the overall volume change is independent of crystallization and amorphization processes at the applied current density of 40 µA cm$^{-2}$ (0.22 C). However, note that for high lithiation rates (e.g., 1.2 C) diffusion-limited lithiation could lead to depth-dependent concentration gradients or structural inhomogeneities, which might change the situation.

Unfortunately, the volume change could not be tracked for a high Li content (x > 3). This is due to the high nSLD of the Cu current collector, which masks the Bragg peak for x > 3.

We would like to emphasize that the measured volume changes of this study are solely due to the modification of the active material germanium. This is because the Bragg peak is due to the modulation of the isotopes inside the volume of the germanium film electrode. Therefore, for the present Bragg peak-based approach, it is not necessary to separate the volume change that appears at the surface (e.g., growth and degrowth of the SEI layer). This is different from NR measurements on single films (see references [7,10], for example). Our NR simulations have revealed that roughening of the surface does not significantly affect the Bragg peak because it is formed by the periodicity of the isotope nSLD within the ML. This is not the case for single films, where only two interfaces exist at which reflections occur. If one of these interfaces, e.g. the surface in contact with the electrolyte, becomes rougher during cycling, the interference fringes in NR are reduced and may diminish with increasing surface roughness.

### 4.4 Further discussion and outlook

For a reliable evaluation of the results, several aspects need to be considered:

(i) A homogeneous lithiation mechanism is required to extract meaningful volume changes. For Ge, this has been demonstrated in ref. [15]. In this context, it should be noted that NR provides only a lateral average over the entire electrode, so local inhomogeneities (e.g., spatially varying insertion behavior) are not resolved.

(ii) The Li content, *x*, is determined from integration of the charge supplied by the potentiostat and is assumed to correspond to the amount of incorporated Li. However, the observed coulombic efficiency (Figure 8b) introduces an uncertainty that directly affects the derived volume–concentration relationship.

(iii) The assumption that SEI effects are fully excluded from the analysis may only be partially valid. While the Bragg peak originates from the IML, parasitic reactions can still influence the amount of Li stored in the Ge IML.

These aspects are discussed in the present section.

A weakness of all NR based approaches to determine volume changes in LIB electrodes is the unknown exact Li content x in the germanium (or silicon) electrode that cannot be determined exactly by the method itself. As previously mentioned, the Li content in germanium was calculated by assuming that the entire charge delivered and obtained by the potentiostat is inserted or extracted as $Li^+$ ions inside the germanium electrode. Side reactions that could consume $Li^+$, such as the formation and shrinking of the SEI layer during cycling are not considered in the calculation of the content x [28,29]. Subtracting the amount of $Li^+$ consumed in the side reactions from the x-values determined, would result in lower x-values for a given volume change (Figures 13 and 14). This will increase the slope of the curve and a higher volume expansion at a given x-value can be expected. Due to the limited thickness of the SEI, this effect is expected to be low. A comparative study of IML electrodes and single film electrode (with SEI expansion) will give more insight in this problem.

Figure 13a shows that the deviation between the measured and predicted values (extrapolated from Si volume change on single films) increases for higher x values. Under the assumption that Si and Ge show the same volume expansion and have the same amounts of free volumes, this may indicate that additionally the SEI grows and not only the volume of the active Ge material.

Up to now, the literature does not provide a clear consensus on the specific potential at which SEI formation occurs (see refs. [28–30]). It is often assumed that SEI growth is favored at relatively high electrode potentials, before significant lithiation of active materials such as carbon, silicon, and germanium begin. In contrast, other studies suggest that SEI formation predominantly takes place at low potentials, where the electrolyte becomes unstable. For carbonate-based electrolytes, for example, it has been shown (ref. [51]) that the formation of a stable SEI at potentials below 1 V is essential for maintaining kinetic stability.

To estimate the impact of SEI formation on the *x*-values shown in Figures 13 and 14, a continuous loss of Li in the IML is assumed, increasing with SOC. This estimation is based on electrochemical data: as shown in Figure 8, the coulombic efficiency ($\eta$) is below 100%, indicating Li consumption by side reactions, likely dominated by SEI formation. Accordingly, a correction is applied using $x_{corrected} = x \cdot (\eta/100)$. Figures 15 and 16 present the resulting volume changes based on the corrected *x*-values. With this correction performed for Figures 15 and 16, the agreement improves significantly, approaching near-perfect agreement between the volume changes obtained from NR (red circles in Figure 15a) and the expected values for Ge derived from volume expansion data of silicon films (blue line in Figure 15a).

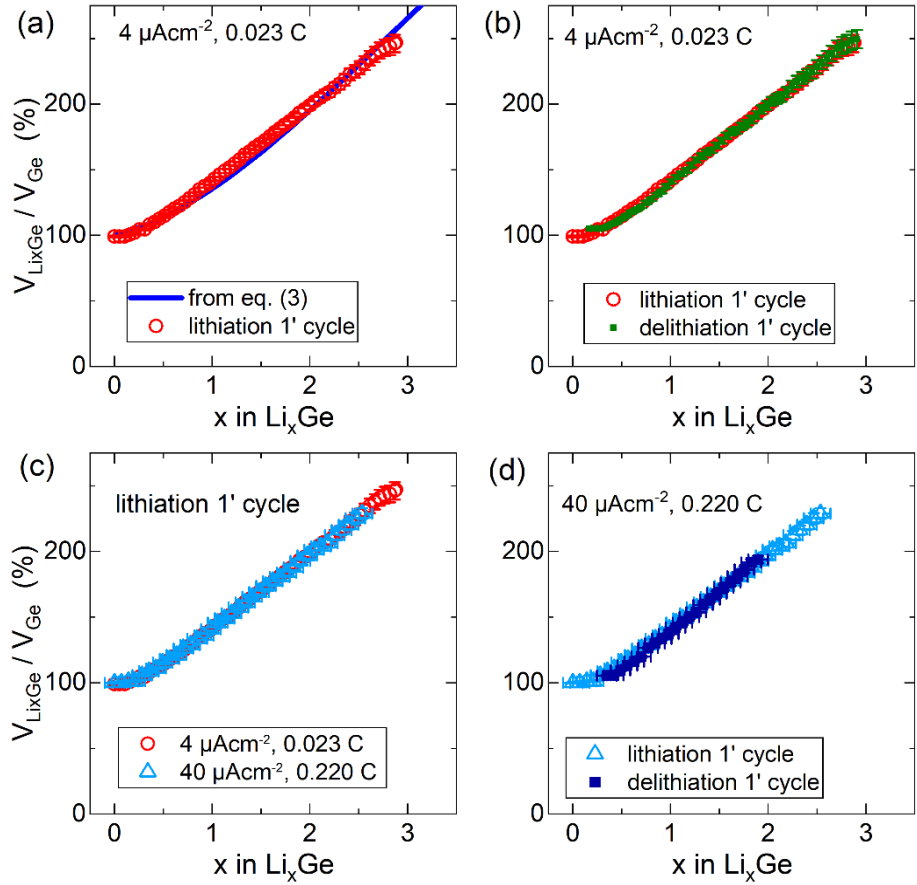

**Figure 15.** Volume changes versus corrected Li content $x = x_{corrected}$ in $Li_xGe$. **(a)** Volume change of the amorphous Ge IML with $d_0 = 7.0$ nm at a C-rate of 0.023 C during first lithiation (red circles) compared to the predicted volume change of amorphous Ge from Si NR data (equation 3). **(b)** Volume change of the IML with $d_0 = 7.0$ nm during first lithiation (red circles) and delithiation (green squares) for the first cycle with a C-rate of 0.023 C. **(c)** Volume change during the first lithiation of the IML with $d_0 = 7.0$ nm with a C-rate of 0.023 C (red circles) and with $d_0 = 6.4$ nm with a C-rate of 0.22 C (blue triangles). **(d)** Volume change of the ML with $d_0 = 6.4$ nm during first lithiation (blue triangles) and delithiation (dark blue squares) at a C-rate of 0.22 C.

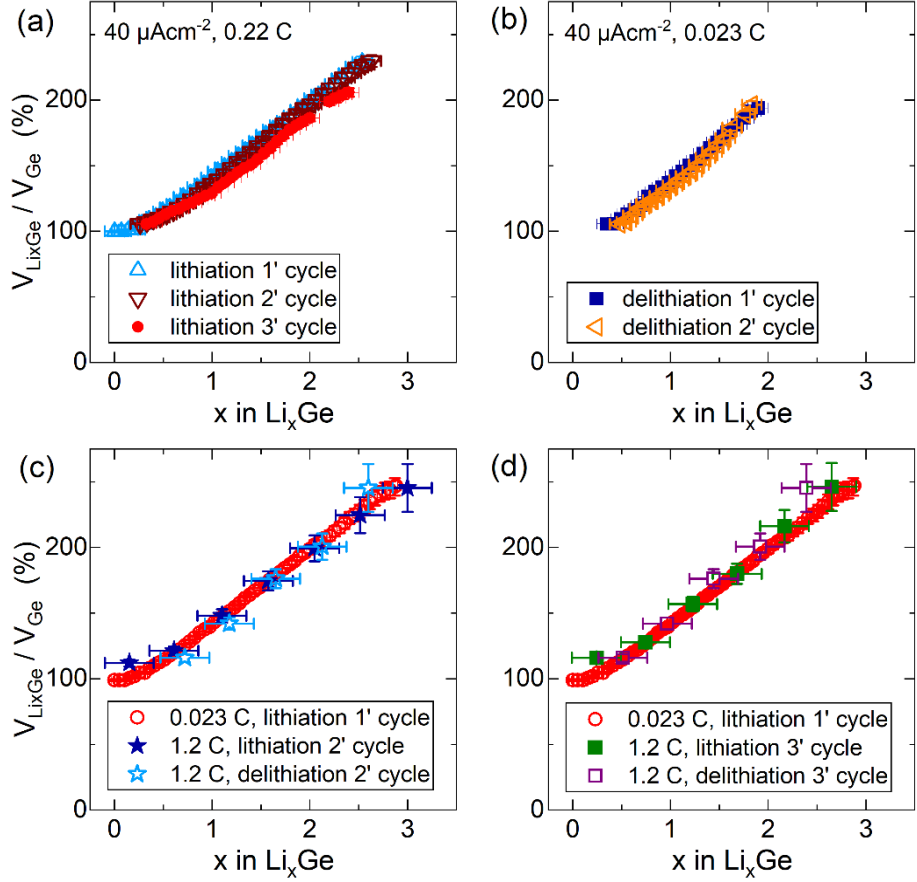

**Figure 16.** Volume changes versus corrected Li content x = $x_{corrected}$ in $Li_xGe$. **(a)** Volume change of the IML with $d_0$ = 6.4 nm during lithiation with 0.22 C for the first (unfilled blue triangles), second (brown triangles), and third (red circles) cycle. **(b)** Volume change of the IML with $d_0$ = 6.4 nm during delithiation with 0.22 C for the first (filled dark blue squares) and the second cycle (orange triangles). **(c)** Volume change of the IML with $d_0$ = 7.0 nm during the second cycle with 1.2 C (dark blue stars for lithiation and unfilled blue stars for delithiation) compared to the first cycle with 0.022 C (unfilled red circles). **(d)** Volume change of the IML with $d_0$ = 7.0 nm during the third cycle with 1.2 C (green squares for lithiation and purple squares for delithiation) compared to the first cycle with 0.022 C (red circles).

In principle, it is possible to determine the real Li content in the Ge IML from the nSLD by NR using Figure 12d. As shown, Li incorporation or extraction from the Ge IML increases or decreases the nSLD. This should result in a decrease or increase in the Bragg peak intensity as was also measured (Figure 9b). Unfortunately, this analysis is not possible, because other process also may lead also to a modification of the Bragg peak intensity, such as plastic deformation. Plastic deformation during cycling [64] can induce structural rearrangements and partial intermixing within the Ge layers. This reduces the isotope contrast of the layered Ge distribution, which in turn leads to a decrease in the neutron Bragg peak associated with the Ge IML structure. Further, delamination of parts of the Ge IML from the current collector may occur, influencing intensity. Consequently, we believe that the intensity variation of the Bragg peak measured during current operando NR measurements cannot be used to determine the Li content in the Ge IML electrode. Therefore, we used equation (5) instead.

The operando NR experiments on germanium IML electrodes performed in the current work suggests that the intrinsic volume change of amorphous germanium films is the same if the material is cycled close to thermodynamic equilibrium (4 µA cm$^{-2}$, 0.023 C), at a moderate cycling rate (40 µA cm$^{-2}$, 0.023 C), and at a fast-cycling rate (153 µA cm$^{-2}$, 1.2 C). These results suggest a high mobility of Li in amorphous germanium electrodes during cycling. Recently, Chou and Hwang [63] investigated the origin of the significant difference in lithiation between crystalline germanium and crystalline silicon using density functional theory calculations. They found that the relatively strong Li-Si interaction and the stiffer silicon lattice tend to reduce Li mobility ($D^{Li} = 10^{-13}$ cm²s$^{-1}$) in crystalline silicon compared to crystalline germanium ($D^{Li} = 10^{-11}$ cm²s$^{-1}$) [65]. With ongoing lithiation, $D^{Li}$ in Li$_x$Si increases significantly, from $10^{-12}$ to $10^{-7}$ cm²s-$^1$ (for x from 0.14 to 3.57). Conversely, $D^{Li}$ in Li$_x$Ge remains around $10^{-7}$ cm²s-$^1$ and is less concentration dependent [65]. Their theoretical analysis shows that rapid Li diffusion in Li$_x$Ge is directly related to facile atomic rearrangements of host Ge atoms, even in the early stages of lithiation. Thus, in addition to Li diffusion, Ge diffusion (i.e. diffusion of the host atom) during cycling of amorphous germanium films also has to be investigated close to room temperature, which may be done also operando with Ge IMLs.

The proof-of-concept for the determination of volume expansion by IML and NR was given in the present work for a liquid electrolyte. However, the IML based method is predestined to work also on all-solid-state batteries. The present experiments were done using specially designed electrochemical cells so that the neutron beam would probe only the single electrode under investigation (the working electrode) and not both electrodes (the counter and working electrode). For this reason, the distance between the positive and negative electrodes was made large enough (in the used cells it was 20 millimeters) that the obtained NR pattern would contain only information from the electrode under investigation (e.g., the germanium electrode) [3-5,8,10,19,20]. Such large distances (a few millimeters) between the electrodes can be achieved for electrochemical cells with liquid electrolytes but not for all-solid-state LIBs. However, using isotope modulated active materials renders this problem irrelevant. The Bragg peak is uniquely determined by the isotope modulation in the IML, allowing to track solely a specific electrode or electrolyte.

It should be noted that single crystalline Ge isotope multilayers have been successfully produced by epitaxy, resulting in a suitable Bragg peak in NR [26]. These MLs can also be used as electrodes in electrochemical cells for operando NR measurements during the lithiation and delithiation cycles of crystalline germanium. The same applies to amorphous silicon [23] and crystalline silicon [27].

The approach of using operando tracking of an NR Bragg peak during cycling is not limited to IMLs. It can also be done for multilayers with a modulation of chemical contrast. Multilayers with silicon or germanium embedded between other materials like carbon will produce a proper Bragg peak. Si/C combinations are desired for use in commercial LIBs nowadays [29]. It is assumed that the carbon matrix, with at least some percentage of silicon embedded within, will relax the volume change of silicon during cycling and restrict it. Therefore, it is interesting to investigate how the volume change in silicon layers is constricted by thin or thick carbon intermediary layers in Si/C or Ge/C MLs. The approach presented in the current study is ideal for such investigations.

## 5. Conclusion

This study presented an advanced method to study volume changes during cycling of LIB electrode materials that is based on IML and operando NR. A proof of concept was carried out on amorphous germanium as the active material in a LIB with a liquid electrolyte. $^{nat}$Ge/$^{73}$Ge multilayer film electrodes were successfully deposited and investigated. The isotope modulation produces a Bragg peak in the NR pattern. The volume change during electrochemical cycling can be deduced from the shift in the scattering vector position of the Bragg peak. Time consuming fitting procedures of numerous complex reflectivity patterns is omitted by using IML. This methodology has the advantage that the volume change is measured exclusively inside the amorphous Ge electrode, independently of battery-side reactions, such as surface roughening, growth, or degradation of a SEI layer.

The reversible volume change was successfully tracked during lithiation and delithiation cycles with constant current densities ranging from close to thermodynamic equilibrium (4 µA cm$^{-2}$, 0.023 C), moderate rate (40 µA cm$^{-2}$, 0.22 C), to fast-rate (153 µA cm$^{-2}$, 1.2 C). The largest observed volume change was 250% for a Li content of x ≈ 3. The obtained results also tentatively indicate that the volume change was found to be independent of (i) cycling speed, (ii) cycle number, (iii) film thickness, and (iv) whether crystallization and subsequent re-amorphization occurred during cycling. Based on this proof of concept, the easy to realize method can be also applied to all-solid-state LIBs and multilayer electrodes with chemical contrast like Si/C.

**Acknowledgements**

This work is based upon experiments performed on AMOR at the Swiss spallation neutron source SINQ, Paul Scherrer Institute, Villigen, Switzerland. This work was funded by the Deutsche Forschungsgemeinschaft (DFG, German Research Foundation) under the contract Schm1569/38-1 (Project number 449439875). The financial support is gratefully acknowledged. We thank Christine Klauser from the Neutron Optics Group of PSI for magnetron sputter deposition of the Cr adhesion layer and of the Cu current collector.

**ORCID**

Erwin Hüger https://orcid.org/0000-0002-1545-1459

Jochen Stahn https://orcid.org/0000-0002-6711-0592

Harald Schmidt https://orcid.org/0000-0001-9389-8507

AUTHOR DECLARATIONS

**Conflict of Interest**

The authors have no conflicts to disclose.

DATA AVAILABILITY

The data that support the findings of this study are available from the corresponding author upon reasonable request.

**Table of Contents (TOC)/Abstract Graphic**

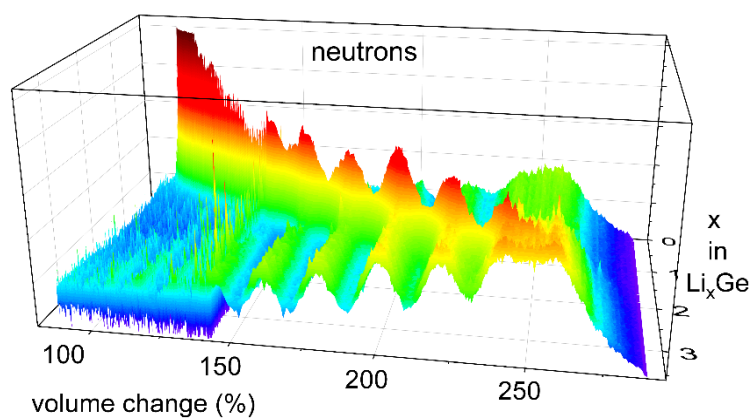